\documentclass[11pt,a4paper]{article}
\usepackage{jheppub}

\pdfoutput=1

\usepackage{comment}

\usepackage{color}
\usepackage{graphicx}
\usepackage{url}
\usepackage[usenames,dvipsnames]{xcolor}
\usepackage{dcolumn}
\usepackage{bm}
\usepackage{bbm}
\usepackage{amsmath,amssymb,amsfonts}
\usepackage{dsfont}
\usepackage[vcentermath]{youngtab}
\usepackage[all]{xy}
\usepackage{pstricks}
\usepackage{dsfont}%
\usepackage{mathtools}
\usepackage{placeins}
\usepackage{enumerate}
\usepackage{cases}
\usepackage{placeins}
\usepackage{xspace}
\usepackage{cancel} 
\usepackage{slashed}
\usepackage{natbib}
\usepackage{subcaption}

\newcommand{\beq}{\begin{eqnarray}}
\newcommand{\eeq}{\end{eqnarray}}

\newcommand{\bmp}{\noindent\begin{minipage}{16cm}}
\newcommand{\emp}{\end{minipage}\vskip 7mm} 

    \newcommand{\ii}{\mathrm{i}}

    \newcommand{\EW}{\mathrm{EW}}
    
    \newcommand{\SU}{\mathrm{SU}} 
    \newcommand{\Sp}{\mathrm{Sp}}
    
    \newcommand{\Tr}{\mathrm{Tr}}

    \newcommand{\SUL}{\mathrm{SU}(2)_{\mathrm{L}}} 
     
        \newcommand{\vw}{v_{\mathrm{w}}} 
    \newcommand{\gL}{g_\mathrm{L}}
    \newcommand{\gY}{g_{Y}}
    \newcommand{\NF}{N_{\mathrm{F}}}
    \newcommand{\RF}{R_{\mathrm{F}}}

    \newcommand{\bee}{\begin{equation}}
        \newcommand{\eee}{\end{equation}}

\def\eq#1{{Eq.~\eqref{#1}}}

\def\fig#1{{Fig.~\ref{#1}}}

\def\fig#1{{Fig.~\ref{#1}}}

\def\lsim{\mathrel{\rlap{\lower4pt\hbox{\hskip1pt$\sim$}}
    \raise1pt\hbox{$<$}}}                
\def\gsim{\mathrel{\rlap{\lower4pt\hbox{\hskip1pt$\sim$}}
    \raise1pt\hbox{$>$}}}                

\begin{document}


\title{Partially composite Higgs models:
\\ 
Phenomenology and RG analysis
}


\author[a]{Tommi Alanne}
\author[b]{Diogo Buarque Franzosi}
\author[c]{Mads T. Frandsen}
\author[c]{Mette L.A. Kristensen}
\author[d]{Aurora Meroni}
\author[c]{Martin Rosenlyst}

\affiliation[a]{Max-Planck-Institut f\"{u}r Kernphysik, Saupfercheckweg 1, 69117 Heidelberg, Germany}
\affiliation[b]{Institut f\"{u}r Theoretische Physik, Universit\"at G\"ottingen,
Friedrich-Hund-Platz 1, 37077 G\"ottingen, Germany}
\affiliation[c]{CP$^{3}$-Origins, University of Southern Denmark, Campusvej 55, DK-5230 Odense M, Denmark}
\affiliation[d]{Department of Physics, University of Helsinki, 
\& Helsinki Institute of Physics, \\
                      P.O.Box 64, FI-00014 University of Helsinki, Finland}

\emailAdd{tommi.alanne@mpi-hd.mpg.de}
\emailAdd{dbuarqu@gwdg.de}
\emailAdd{frandsen@cp3.sdu.dk}
\emailAdd{metkr03@student.sdu.dk}
\emailAdd{aurora.meroni@helsinki.fi}
\emailAdd{rosenlyst@cp3.sdu.dk}

\abstract{

We study the phenomenology of partially composite-Higgs models where electroweak symmetry breaking is dynamically induced, and the
Higgs is a mixture of a composite and an elementary state. 
The models considered have explicit realizations in terms of gauge-Yukawa theories 
with new strongly interacting fermions coupled to elementary scalars and allow for a very SM-like Higgs state. 
We study constraints on their parameter spaces from vacuum stability and perturbativity as well as from LHC results
and find that requiring vacuum stability up to the compositeness scale already imposes relevant constraints. A small part of parameter space around the classically conformal limit is stable up to the Planck scale. This is however already strongly disfavored by LHC results.
In different limits, the models
realize both (partially) composite-Higgs and (bosonic) technicolor models and a dynamical
extension of the fundamental Goldstone-Higgs model. 
Therefore, they provide a general framework for exploring the phenomenology of composite dynamics.

{\footnotesize  \it Preprint: CP3-Origins-2017-054 DNRF90}
}

\maketitle
\newpage

\section{Introduction}
\label{sec:intro}

Partially composite-Higgs (CH) models with underlying four-dimensional realizations in terms of new fermions coupled to elementary scalar multiplets
via Yukawa interactions~\cite{Kaplan:1983fs} were recently developed and studied in 
Refs.~\cite{Galloway:2016fuo,Agugliaro:2016clv,Alanne:2016rpe,Alanne:2017rrs}.
These models feature dynamically induced electroweak symmetry breaking (EWSB) and a partially composite
pseudo-Goldstone boson (pGB) Higgs. The elementary scalar multiplet provides masses for the SM fermions via SM-like Yukawa
interactions \cite{'tHooft:1979bh} avoiding the complications of fermion partial compositeness~\cite{Kaplan:1991dc} or 
extended-technicolor-type 
constructions~\cite{Eichten:1979ah,Dimopoulos:1979es,Kaplan:1983sm}.
The framework gives a unified description 
for various classes of models from (partially) CH~\cite{Kaplan:1983fs} to (bosonic) technicolor (b)TC~\cite{Simmons:1988fu,Kagan:1990az}. 

However, such models may suffer from a low 
vacuum instability scale due to the enhancement of the top Yukawa coupling of the Higgs state as compared to the SM Higgs~\cite{Carone:2012cd}. 
In this paper
we, therefore, study the vacuum stability and perturbativity bounds on the parameter space of minimal partially CH (pCH) 
models~\cite{Galloway:2016fuo,Agugliaro:2016clv,Alanne:2017rrs}
and combine these self-consistency conditions with current constraints
from collider experiments.
We show that even the requirement of stability up to the compositeness scale provides relevant constraints regardless of the particular UV realization. 
We also consider model extensions where a (small) part of the top-quark mass does not originate from the vacuum expectation value (vev) of the elementary scalar multiplet 
as a possibility to alleviate the vacuum-stability bounds. 

The two underlying concrete models we consider both consist 
of four Weyl fermions, $U_{L,R}, D_{L,R}$, transforming under a new $\SU(2)_{\mathrm{TC}}$ gauge group with the 
left-handed components forming a doublet, $(U_{\mathrm{L}},D_{\mathrm{L}})$, under the weak gauge group $\SU(2)_{\mathrm{w}}$. This 
is the minimal CH model with four-dimensional fermionic realisation \cite{Ferretti:2013kya}, and the new fermion sector features an enhanced $\SU(4)$
global symmetry acting on the four Weyl fermions. The fermions are 
coupled to a multiplet of new elementary scalars via Yukawa interactions, and the dynamical condensation of the strongly interacting
(techni)fermions is transmitted to the scalar multiplet via these Yukawa interactions.  As a consequence, the 
fundamental scalar multiplet acquires a vev. We begin by considering a $\SU(2)_{\mathrm{w}}$ doublet of 
scalars~\cite{Galloway:2016fuo,Agugliaro:2016clv} and then study 
a full multiplet under the global $\SU(4)$ symmetry of 
the new fermion sector~\cite{Alanne:2017rrs}. 

While the SM fermions obtain their masses via the vev of the elementary scalar multiplet, the weak boson masses
originate from both the vev and the condensate in the composite sector. Consequently, the electroweak (EW) scale is given by
\begin{equation}
    \label{eq:vevs}
    \vw^2=v^2+f^2\sin^2\theta,  
\end{equation}
where $\vw=246$~GeV, we denote the vev  of the neutral CP-even component of the elementary Higgs by $v$,
$f$ is the Goldstone-boson decay constant of the composite sector, and $\theta$ ($ \pi/2\leq \theta \leq \pi$) parameterises the 
vacuum misalignment.

\section{Minimal partially composite Higgs models}
\label{sec:elemComp}

We consider here the minimal four-dimensional strongly interacting fermionic sector allowing for a composite Goldstone Higgs boson. 
This is based on the gauge group $\SU(2)_{\mathrm{TC}}$ and contains four Weyl fermions
transforming in the fundamental representation of the gauge group. The new fermion sector in isolation features a global $\SU(4)$ symmetry,
which upon condensation breaks spontaneously to $\Sp(4)$ \cite{Lewis:2011zb}. The left-handed fermions $(U_L,D_L)$ transform as a doublet under $\SU(2)_{\mathrm{w}}$,
and the EW gauge group can be embedded into the $\SU(4)$ global symmetry
via the left and right generators
\begin{equation}
    \label{eq:gensLR}
    T^i_{\mathrm{L}}=\frac{1}{2}\left(\begin{array}{cc}\sigma_i & 0 \\ 0 & 0\end{array}\right),\
    T^i_{\mathrm{R}}=\frac{1}{2}\left(\begin{array}{cc} 0 & 0 \\ 0 & -\sigma_i^{T}\end{array}\right).
\end{equation}
The $T_{\mathrm{L}}$ generators may be identified with those of the $\SU(2)_{\mathrm{w}}$ and $T_{\mathrm{R}}^3$  
with the generator of hypercharge~\cite{Luty:2004ye,Galloway:2010bp,Cacciapaglia:2014uja,Franzosi:2016aoo}.
The five Goldstone bosons (GBs) associated to the breaking $\SU(4)\rightarrow\Sp(4)$ then form an EW doublet and a singlet.
We list the left-handed fermions and their 
quantum numbers using the notation $\widetilde{U}_{\mathrm{L}}\equiv \epsilon U_R^*$ in Tab.~\ref{tab:su4}. The details of the 
construction of the effective description can be found e.g. in Refs.~\cite{Galloway:2016fuo,Alanne:2017rrs}. 
\begin{table}[h!]
    \caption{New fermion content.}
    \label{tab:su4}
    \begin{center}
	\begin{tabular}{cccc}
	    \hline
	    & $\vphantom{\frac{\frac12}{\frac12}}\quad\SU(2)_{\mathrm{TC}}\quad$ & $\SU(2)_{\mathrm{W}}\quad$ 
	    & $\mathrm{U}(1)_Y\quad$\\
	    \hline
	    $\vphantom{\frac{\frac12}{\frac12}} Q_{\mathrm{L}}$	&   ${\tiny \yng(1)}$	&   ${\tiny \yng(1)}$	&   0\\    
	    $\vphantom{\frac{1}{\frac12}} \widetilde{U}_{\mathrm{L}}$	&   ${\tiny \yng(1)}$	&   1	&   $-1/2$\\    
	    $\vphantom{\frac{1}{\frac12}} \widetilde{D}_{\mathrm{L}}$	&   ${\tiny \yng(1)}$	&   1	&   $+1/2$ \\
	    \hline
	\end{tabular}
    \end{center}
\end{table}

There are different possibilities to align the stability group, $\Sp(4)$, with respect to the 
EW gauge symmetry inside the $\SU(4)$. 
In particular, the EW group can be embedded completely within the stability group leading to unbroken EW symmetry. 
There are two inequivalent vacua, $E_{\pm}$, corresponding to this orientation.
On the other hand, it is possible that only the electromagnetic subgroup, $\mathrm{U}(1)_Q$, is unbroken as in TC models~\cite{Weinberg:1975gm,Susskind:1978ms}.
We denote the corresponding vacuum by $E_{\mathrm{B}}$. 
The different vacua can be represented as the matrices
\begin{equation}
    E_\pm = \left( \begin{array}{cc}
	\ii \sigma_2 & 0 \\
	0 & \pm \ii \sigma_2
    \end{array} \right)\,,\qquad
    E_B  =\left( \begin{array}{cc}
	0 & 1 \\
	-1 & 0
    \end{array} \right).
\end{equation}
In general, the vacuum is a superposition of an EW-preserving and the EW-breaking vacua, and can be written as
\begin{equation}
E=\cos\theta E_-+\sin\theta E_{\mathrm{B}}. 
\end{equation}
The misalignment angle, $\theta$, is determined by the sources of explicit SU(4) breaking, and the relation between the composite scale 
and the EW scale is then given in terms of  $\theta$ according to 
Eq.~\eqref{eq:vevs}.

We parameterise the composite Goldstone-boson degrees of freedom in the $\SU(4)/\Sp(4)$ 
coset by
\begin{equation}
    \label{eq:}
    \Sigma=\exp\left(\frac{2\sqrt{2}\,\ii}{f}\Pi^a X^a\right)E,
\end{equation}
where the broken generators, $X^a$, with $a=1,\dots,5$, correspond to the vacuum $E$. The generators are explicitly given in Ref.~\cite{Galloway:2010bp}.
Furthermore, we introduce spurions, $P_{\alpha}$, $\widetilde{P}_{\alpha}$,
such that $\Tr[P_{\alpha}\Sigma]$ and $\Tr[\widetilde{P}_{\alpha}\Sigma]$ transform as EW doublets with hypercharges $+1/2$
and $-1/2$, respectively. These can be explicitly written as
\begin{equation}
    \label{eq:proj}
    \begin{split}
	2P_{1}&=\delta_{i1}\delta_{j3}-\delta_{i3}\delta_{j1},\ 
	2P_{2}=\delta_{i2}\delta_{j3}-\delta_{i3}\delta_{j2},\\
	2\widetilde{P}_{1}&=\delta_{i1}\delta_{j4}-\delta_{i4}\delta_{j1},\ 
	2\widetilde{P}_{2}=\delta_{i2}\delta_{j4}-\delta_{i4}\delta_{j2}.
    \end{split}
\end{equation}

    \subsection{Model I: Elementary scalar doublet}
    \label{sec: Elementary scalar doublet}

    In addition to the composite pions, $\Sigma$, we introduce an (elementary) scalar multiplet. We begin by considering an $\SUL$ doublet,
$H_\alpha$, and we also introduce vector-like masses for the new fermions~\cite{Galloway:2016fuo,Agugliaro:2016clv} via the matrix $M$: 
    \begin{equation}
	\label{eq:H}
	H_\alpha=\frac{1}{\sqrt{2}}\left(\begin{array}{c}\sigma_h-\ii \pi_h^3 \\ -(\pi_h^2+\ii \pi_h^1)\end{array}\right) , 
	    \quad M=\left(\begin{array}{cc}m_1\epsilon & 0\\ 0 & -m_2\epsilon\end{array}\right) .
    \end{equation}
    With this field content, the 
    underlying Lagrangian describing the new strong sector and the elementary doublet can be written as 
    \begin{equation}
	\label{eq:UVLag1}
	\begin{split}
	    \mathcal{L}_{\mathrm{pCH}}=&\bar{Q}\ii\slashed{D}Q+D_{\mu}H^{\dagger}D^{\mu}H-m_H^2H^{\dagger}H-\lambda_H(H^{\dagger}H)^2\\
	    &+\frac{1}{2}Q^T\,M\,Q- y_U H_{\alpha}(Q^{T}P_{\alpha}Q)-y_D \widetilde{H}_{\alpha}(Q^{T}\widetilde{P}_{\alpha}Q)+\mathrm{h.c.},
	\end{split}
    \end{equation}
        where  
    $Q=(U_{\mathrm{L}},\, D_{\mathrm{L}},\, \widetilde{U}_{\mathrm{L}},\, \widetilde{D}_{\mathrm{L}}\,)$, 
    $\widetilde{H}=\epsilon H^*$, and the antisymmetric contractions
    are kept implicit. 
    The global $\SU(4)$ symmetry of the composite sector is in this case broken explicitly to $\SU(2)_{\mathrm{w}}\times \mathrm{U}(1)_Y$ by 
    both the Yukawa and weak interactions, and the vector like masses in $M$.
    The Yukawa couplings with the SM fermions, restricting to the third generation of SM quarks,  
     are 
    \begin{equation}
	\label{eq:}
	\mathcal{L}_{\mathrm{Yuk}}=- y_t \bar{q}_{\mathrm{L}} H t_{\mathrm{R}}- y_b \bar{q}_{\mathrm{L}} \widetilde{H} b_{\mathrm{R}}+ \text{ h.c.},
    \end{equation}
    where $q_L=(t_L,b_L)$.
    This model was considered in Refs.~\cite{Galloway:2016fuo,Agugliaro:2016clv} in the limit $\lambda_H=0$ but here we keep $\lambda_H$ throughout,
    since it is relevant for the vacuum 
    stability, and since we do not assume necessarily any supersymmetric completion that would alter the running of $\lambda_H$ at high energies. 

    Below the condensation scale, $\Lambda_{\rm TC} \sim 4\pi f$, Eq.~\eqref{eq:UVLag1} yields the following lowest-order effective potential
    \begin{equation}
	\label{eq:effpot}
	\begin{split}
	    V_{\mathrm{eff}}^0=&4\pi f^3Z_2\left(\vphantom{\widetilde{H}_{\alpha}}\frac{1}{2}\Tr\left[M\Sigma\right]
		+y_U H_\alpha\Tr\left[P_{\alpha}\Sigma\right]+y_D\widetilde{H}_{\alpha}\Tr[\widetilde{P}_{\alpha}\Sigma]+\ 
		\text{h.c.}\right)\\
	    &+m_H^2H^{\dagger}H+\lambda_H(H^{\dagger}H)^2,
	\end{split}
    \end{equation}
    where $Z_2$ is a non-perturbative $\mathcal{O}(1)$ constant, and we use the numerical value $Z_2\approx 1.5$ suggested by the 
    lattice simulations \footnote{Ref.~\cite{Arthur:2016dir} gives $\langle Q^TQ\rangle^{1/3}/f=4.19(26)$
    for the condensate, which then yields an estimate (with a factor 4 accounting for the trace):
    $Z_2=\frac{1}{4\cdot 4\pi}\left(\frac{\langle Q^TQ\rangle^{1/3}}{f}\right )^3\approx 1.5$.}~\cite{Arthur:2016dir}.
    The EW gauge interactions contribute to the effective potential at the one-loop level,
    but the contribution is subleading as compared to the vector-like mass terms.  It is, however, important for the model we describe 
    in the next section.

    \subsection{Model II: SU(4) completion of the scalar sector}
    \label{sec:modelII}

    A model retaining the global $\SU(4)$ symmetry on the new fermion sector in the presence of the Yukawa interactions 
    was proposed in Ref.~\cite{Alanne:2017rrs}. The elementary scalar doublet is extended to a complete two-index antisymmetric $\SU(4)$ 
    representation, $\Phi$,
    allowing
    for the Yukawa interactions of the elementary scalars and the new fermions to remain $\SU(4)$ symmetric. Furthermore, 
    the vector-like masses, $m_{1,2}$, of the
    new fermions above are then generated via the dynamically induced vev of the EW-singlet component of the scalar multiplet, $\Phi$. 
    The misalignment of the vacuum, $\theta\neq 0$, required to give a mass to the would-be pGB Higgs boson may come solely 
    from the SM sector and the scalar 
    potential.
    In the following, we briefly outline the model and the construction of the effective low-energy Lagrangian, but we refer to 
    Ref.~\cite{Alanne:2017rrs} for details. 
   
    The scalar multiplet, $\Phi$, can be conveniently parameterised in terms of the EW eigenstates. 
    To this end, we write 
    $H_\alpha \equiv \Tr[P_{\alpha}\Phi]$ and $\widetilde{H}_\alpha\equiv \Tr[\widetilde{P}_{\alpha}\Phi]$ that transform as EW doublets with 
    hypercharges $+1/2$ and $-1/2$, respectively. 
    The EW-singlet directions can be projected out with the following spurions:
    \begin{equation}
	\label{eq:}
	P^{\mathrm{S}}_1=\frac{1}{2}\left(\begin{array}{cc}-\ii\sigma_2 & 0\\ 0 & 0\end{array}\right),\ 
	P^{\mathrm{S}}_2=\frac{1}{2}\left(\begin{array}{cc}0 & 0\\ 0 & \ii\sigma_2\end{array}\right).
    \end{equation}
    Thus, we arrive at
    \begin{equation}
	\label{eq:Phidef}
	\Phi=\sum_{\alpha=1,2}P_{\alpha}H_\alpha+\widetilde{P}_{\alpha}\widetilde{H}_{\alpha}
	    +P^{\mathrm{S}}_1 S+P^{\mathrm{S}}_2 S^{*},
    \end{equation}
    where 
    $H$ is given in Eq.~\eqref{eq:H}, and $S=\frac{1}{\sqrt{2}}(S_{\mathrm{R}}+\ii S_{\mathrm{I}})$ parameterises the EW-singlet scalars. 
    The relevant 
    Lagrangian, restricting to 
    third generation of SM quarks, is then 
    \begin{equation}
	\label{eq:UVLag}
	\begin{split}
	    \mathcal{L}_{\mathrm{pCH}_2}=&\bar{Q}\ii\slashed{D}Q+ \mathrm{Tr}[D_{\mu}\Phi^{\dagger}D^{\mu}\Phi]
		-m_\Phi^2 \mathrm{Tr}[\Phi^{\dagger}\Phi]-\lambda_\Phi \mathrm{Tr}[\Phi^{\dagger}\Phi]^2\\
	    &-y_QQ^T\Phi Q - y_t \bar{q}_{\mathrm{L}} H t_{\mathrm{R}}- y_b \bar{q}_{\mathrm{L}} \widetilde{H}b_{\mathrm{R}}+\mathrm{h.c.}
	\end{split}
    \end{equation}

    Below the condensation scale, 
    we have the lowest-order effective potential
    \begin{equation}
	\label{eq:VeffSU4}
	\begin{split}
	V_{\mathrm{eff}}^0=&4\pi f^3Z_2\left(y_Q \Tr\left[\Phi\Sigma\right]+\ \text{h.c.}\right)
	+m_{\Phi}^2\Tr[\Phi^{\dagger}\Phi]+\lambda_{\Phi}\Tr[\Phi^{\dagger}\Phi]^2.
	\end{split}
    \end{equation}
    At this level, the global symmetry is broken only spontaneously by the condensation in the strong sector. Therefore, in 
    order to misalign the vacuum with respect to EW symmetry and to give a mass to one of the GB's to be identified with the Higgs, 
    we need to consider sources that break the global symmetry explicitly.
    To this end, we include the leading EW contribution in the effective potential~\cite{Peskin:1980gc,Preskill:1980mz}
    \begin{equation}
	\label{eq:EWgauge}
	\begin{split}
	V_{\mathrm{gauge}}=&-C_g\left [ g^2f^4\sum_{i=1}^3 \mathrm{Tr}\left(T_{\mathrm{L}}^i\Sigma 
	    (T_{\mathrm{L}}^i\Sigma)^*\right)+g^{\prime\,2}f^4 \mathrm{Tr}\left(T_{\mathrm{R}}^3\Sigma
	    (T_{\mathrm{R}}^3\Sigma)^*\right)\right ],
	\end{split}
    \end{equation}
    where $C_g$ is a positive $\mathcal{O}(1)$ loop factor.
    Differently from Ref.~\cite{Cacciapaglia:2014uja}, the top quark interactions are subleading in the vacuum determination, 
    since the top quark only couples to the weakly interacting fundamental sector.
    We add another source of explicit breaking of the global symmetry 
    by splitting of the masses of the EW-doublet and singlet 
    components of the scalar multiplet, $\Phi$: 
    \begin{equation}
	\label{eq:VDeltam}
	\begin{split}
	    V_{\delta m^2}=&2\delta m^2\, \Tr[P_i^S\Phi]\Tr[P_i^S\Phi]^*=\frac{1}{2}\delta m^2(S_{\mathrm{R}}^2+S_{\mathrm{I}}^2).
	\end{split}
    \end{equation}
    As shown in Ref.~\cite{Alanne:2017rrs}, these two sources are enough to achieve a non-trivial vacuum alignment, and therefore
    we need not introduce further explicit breaking sources directly on the new fermion sector.

\section{Model I: RG analysis and phenomenological constraints}

    \subsection{Vacuum, spectrum, and $\lambda_H$}

    The effective potential of Eq.~\eqref{eq:effpot} as a function of $\sigma_h$ and $\theta$ reads 
    \begin{equation}
	\label{eq:}
	\begin{split}
	    V_{\mathrm{eff}}=&8\pi f^3Z_2\left(m_{12}c_{\theta}-\frac{y_{UD}}{\sqrt{2}}\sigma_h s_{\theta}\right)
	    +\frac{1}{2}m_H^2\sigma_h^2+\frac{\lambda_H}{4}\sigma_h^4,
	\end{split}
    \end{equation}
    where $m_{12}\equiv m_1+m_2$ and $y_{UD}\equiv y_U+y_D$. We use the short-hand notations $s_x\equiv \sin x$, 
    $c_x\equiv \cos x$, and $t_x\equiv \tan x$.
    Further, we define the parameter $m^2_\lambda \equiv m_H^2 + \lambda_H v^2 $, the form factor $F_0=4\pi Z_2$ 
    and trade $v$ for the angle $\beta$ via
    \begin{equation}
	\label{eq:tbeta}
	t_{\beta}\equiv\frac{v}{f s_{\theta}} . 
    \end{equation} 
    The vacuum conditions then read
    \begin{equation}
	\label{eq:vacCondI}
	\begin{split}
	    0=\left.\frac{\partial V_{\mathrm{eff}}}{\partial \sigma_h}\right|_{\sigma_h=v}=&
		-\sqrt{2}F_0 y_{UD}f^3s_{\theta}+m^2_\lambda v,\\
	    0=\left.\frac{\partial V_{\mathrm{eff}}}{\partial \theta}\right|_{\sigma_h=v}=&
		-2F_0f^3\left(m_{12}s_{\theta} + \frac{y_{UD}}{\sqrt{2}}vc_{\theta}\right).
	\end{split}
    \end{equation}
    These yield the parameter constraint equations
    \begin{equation}
	\label{eq:vac1}
	    y_{UD}= \frac{t_\beta m^2_\lambda}{ \sqrt{2} f^2 F_0 },\quad 
	    m_{12}= - \frac{c_{\theta} m^2_\lambda t_\beta^2 }{2 F_0 f }.
    \end{equation}
    For $t_\beta \gtrsim 1$, the first equation limits the mass parameter 
    to $m_H^2\lesssim 4\pi f^2$ as long as $y_{UD}\lesssim 1$. 
    If the latter limit on $y_{UD}$ is not satisfied, then these large Yukawa couplings will drive the scalar quartic negative at very low scales. 
    Indeed we find that $y_{UD}\ll 1$ for all the viable parameter space that we consider, and this condition is thus 
    automatically satisfied. 
    
    The model features seven parameters relevant to our study, \footnote{Only the combinations $m_{12}$ and $y_{UD}$ enter into the vacuum 
    equations and the effect of $y_U,y_D$ on the running of the couplings is negligible in the parameter space considered.} 
    $ y_{UD}$, $m_{12}$, $\lambda_H$, $s_\theta$, $t_\beta$, $f$, $m_H^2,$ 
    and four constraint equations including the 
    two vacuum conditions in Eq.~\eqref{eq:vacCondI}, the EW scale Eq.~\eqref{eq:vevs}, and the Higgs mass, whose expression is shown below in
    Eq.~\eqref{Eq:masses}. 
    We take $s_\theta, t_\beta$ and $m_H^2$ as free parameters.

    The Yukawa coupling between the top quark and the elementary doublet is enhanced with respect to the SM value by
    \begin{equation}
	\label{eq:topyukawa}
	y_t=y_t^{\mathrm{SM}}/s_{\beta}.
    \end{equation}
    For a given value of the quartic coupling, $\lambda_H$, in Eq.~(\ref{eq:UVLag1}), this implies a lower limit on $t_{\beta}$ in order to 
    avoid vacuum instability above the EWSB scale and below the Planck scale. We discuss this issue in detail in Sec.~\ref{sec:RG}. 
    
    The charged pion and neutral CP-even scalar mass matrices, in the bases  $(\pi_h^+,\Pi^+)$, $(\sigma_h,\Pi^4)$, resp., can be written as
    \begin{equation}
	M_{\pi^+}^2
	    \hspace{-0.1cm}=\hspace{-0.1cm} m^2_\lambda  \left(\begin{array}{cc}
		1 & t_{\beta}\\
		 \hspace{-0.1cm}t_{\beta} &
		 t_{\beta}^2
	    \end{array}\hspace{-0.15cm}\right),
	\quad
	M_{h}^2
	    \hspace{-0.1cm}=\hspace{-0.1cm} m^2_\lambda \left(\begin{array}{cc}
		1+ \delta  & \hspace{-0.05cm}-c_{\theta}t_{\beta} \\
		 \hspace{-0.15cm}-c_{\theta}t_{\beta} &
		 t_{\beta}^2
	    \end{array}\hspace{-0.15cm}\right),
	\label{Eq:masseigenstates}
    \end{equation}	
    where $\delta \equiv 2 \frac{\lambda_H v^2}{m^2_\lambda} $. The mass of the heavy neutral pion, $\pi^0$, coincides with
    the mass of the charged ones.
 
    The CP-even mass eigenstates are given in terms of the interaction eigenstates by
    \begin{equation}
	h_1 = c_\alpha \sigma_h - s_\alpha  \Pi^4,\quad
	h_2 = s_\alpha \sigma_h + c_\alpha  \Pi^4,
    \end{equation}    
    with
    \begin{equation}
	t_{2\alpha}=\frac{2t_\beta c_\theta}{1-t_\beta^2+\delta}\,.
	\label{eq:tan2alpha}
    \end{equation} 
    The corresponding masses of the neutral and charged eigenstates are 
    \begin{equation}
	\label{Eq:masses}
	\begin{split}
	    m^2_{h_{1,2}}=&\frac{m_\lambda^2}{2} \left[1+ t_\beta^2+ \delta \pm \left( 2c_{\theta} t_{\beta} s_{2\alpha}+(1-t_{\beta}^2+\delta)c_{2\alpha}\right)\right] , \\
		m_{\pi^{\pm,\,0}}^2=& m^2_\lambda/c_{\beta}^2,\quad m_{\Pi_5}^2 = t_\beta^2 m_\lambda^2.
	\end{split}
    \end{equation}
    Using Eq.~\eqref{eq:tan2alpha} and expanding in $s_\theta^2$ and  $t_\beta^{-2}$, we then find
    \begin{equation}
	\label{eq:lighthiggs}
	m_{h_1}^2\simeq m_\lambda^2 \left(s_\theta^2+(1-t_{\beta}^{-2})\delta\right),  
	\quad m_{h_2}^2\simeq m_{\lambda}^2 \left(1+ t_\beta^2\right),
	\quad m_{\pi^{\pm,0}}^2= m^2_\lambda \left(1+t_{\beta}^2\right).
    \end{equation}
    
    We now identify the lightest of the neutral CP-even eigenstates with 
    the 125-GeV Higgs boson. In most of parameter space this is the $h_1$ state, but for  $1+\delta > t_\beta^2$ it is instead 
    $h_2$ as seen from Eq.~(\ref{Eq:masseigenstates}). We denote the heavier CP-even eigenstate by ${\cal H}$.

\begin{figure}
	\begin{center}
	    \includegraphics[width=0.45\textwidth]{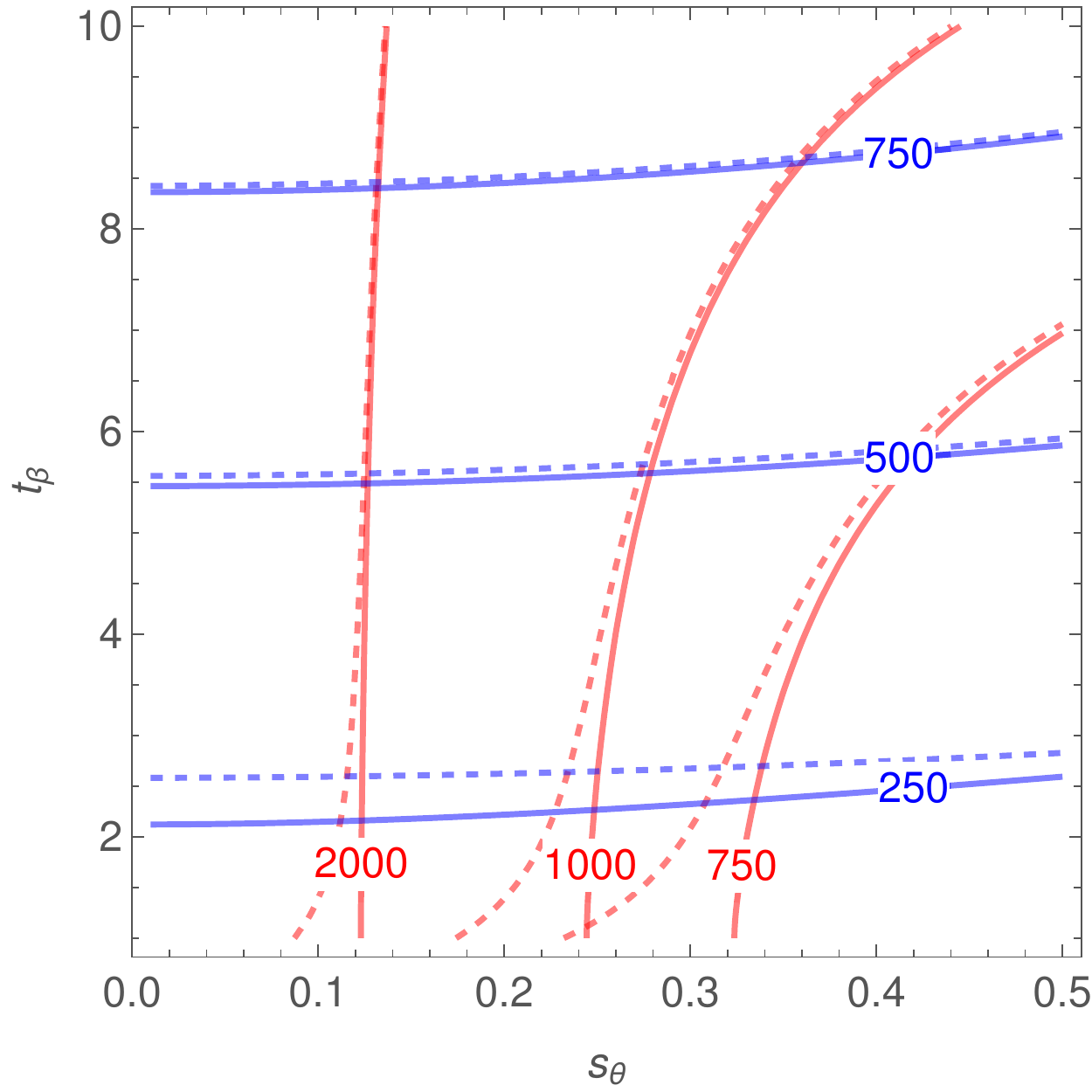}
	     \includegraphics[width=0.45\textwidth]{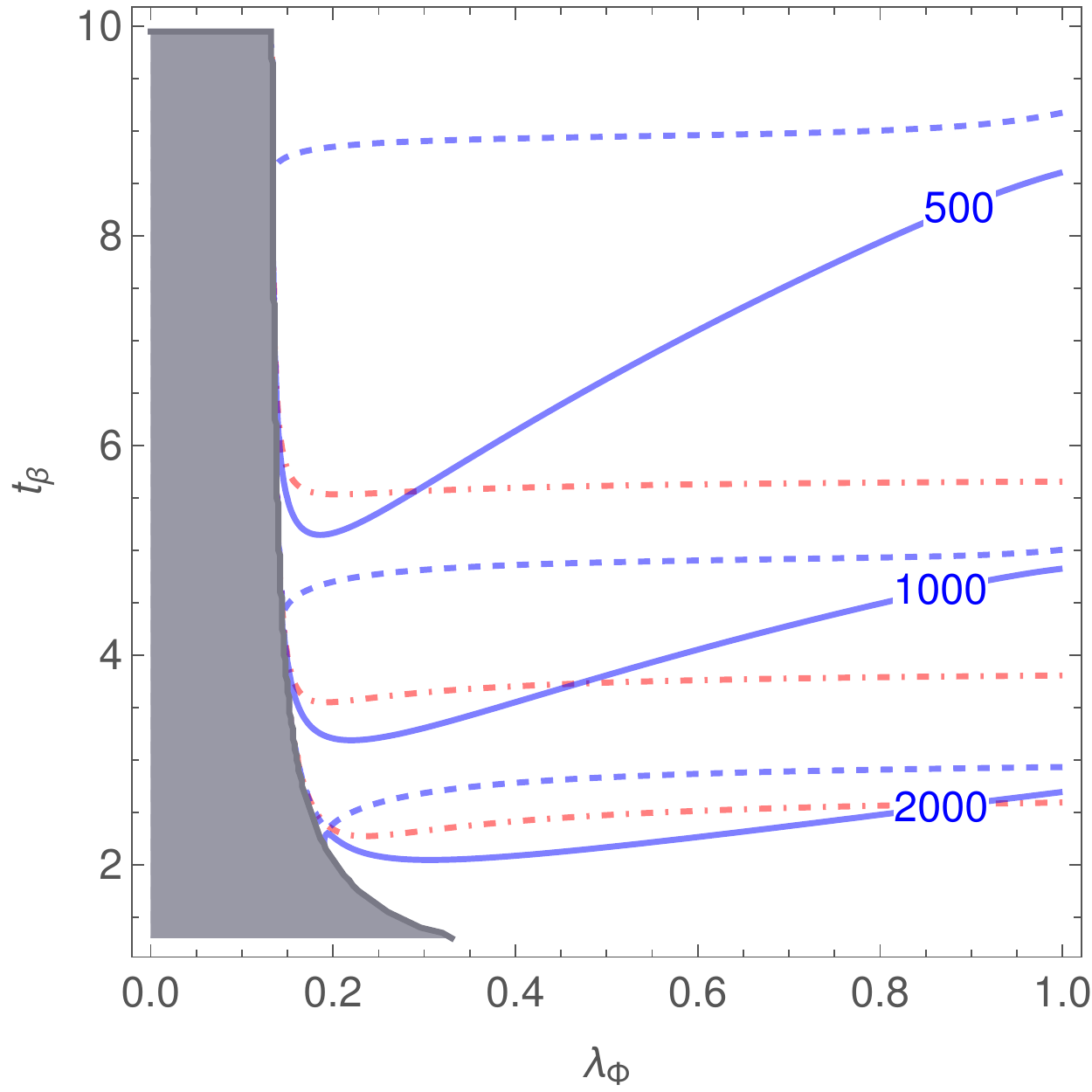}
	\end{center}
	\caption{{\bf Left panel:} Model I spectrum. The blue (red) solid contours show the mass of the heavier CP-even state, ${\cal H}$, in GeV for fixed 
	$m_H^2/f^2=0$ ($m_H^2/f^2=1$). The dashed curves show the mass of the pseudoscalar $\Pi_5$ for the same scenarios.
	{\bf Right panel:}  Model II spectrum. The solid (dashed) blue lines show
	the mass contours (500 GeV, 1000 GeV, 2000 GeV) of $m_{h_2}$ ($m_{h_3}$). The mass of the heavy pion triplet follows
	closely the contours of $m_{h_3}$ , and we have omitted those for clarity. The dot-dashed red lines show the
	same mass contours for the non-pGB mass eigenstate of the two mixing pseudoscalar states, $\Pi_5$ and $S_{\mathrm{I}}$. 
	We have fixed $s_{\theta}= 0.1$.}
	\label{fig:spectrum}
    \end{figure}

    We show the mass of ${\cal H}$ for 
    fixed $m_H^2/f^2=0,1$ in the $(s_{\theta},t_{\beta})$ plane in the left panel of Fig~\ref{fig:spectrum}. The simple scaling of the heavy masses with $t_\beta$, 
    roughly independent of $s_\theta$, is apparent in the $m_H^2/f^2=0$ case  (blue lines).
    Instead the dependence on  $s_\theta$ in the case $m_H^2/f^2=1$ (red dotted lines) follows from the vev condition Eq.~(\ref{eq:vevs}) 
    implying the scaling $m_\lambda^2\simeq m_H^2\simeq v_\EW^2 s_\theta^{-2} t_\beta^{-2}$ which cancels the $t_\beta^2$ factor 
    in the heavy mass formulas. The
    mass of the heavy pion triplet nearly coincides with the mass of the heavier CP-even state for the parameters we consider, 
    and we have omitted that for clarity in the plot.
    
\begin{figure}
	\begin{center}
	  \includegraphics[width=0.45\textwidth]{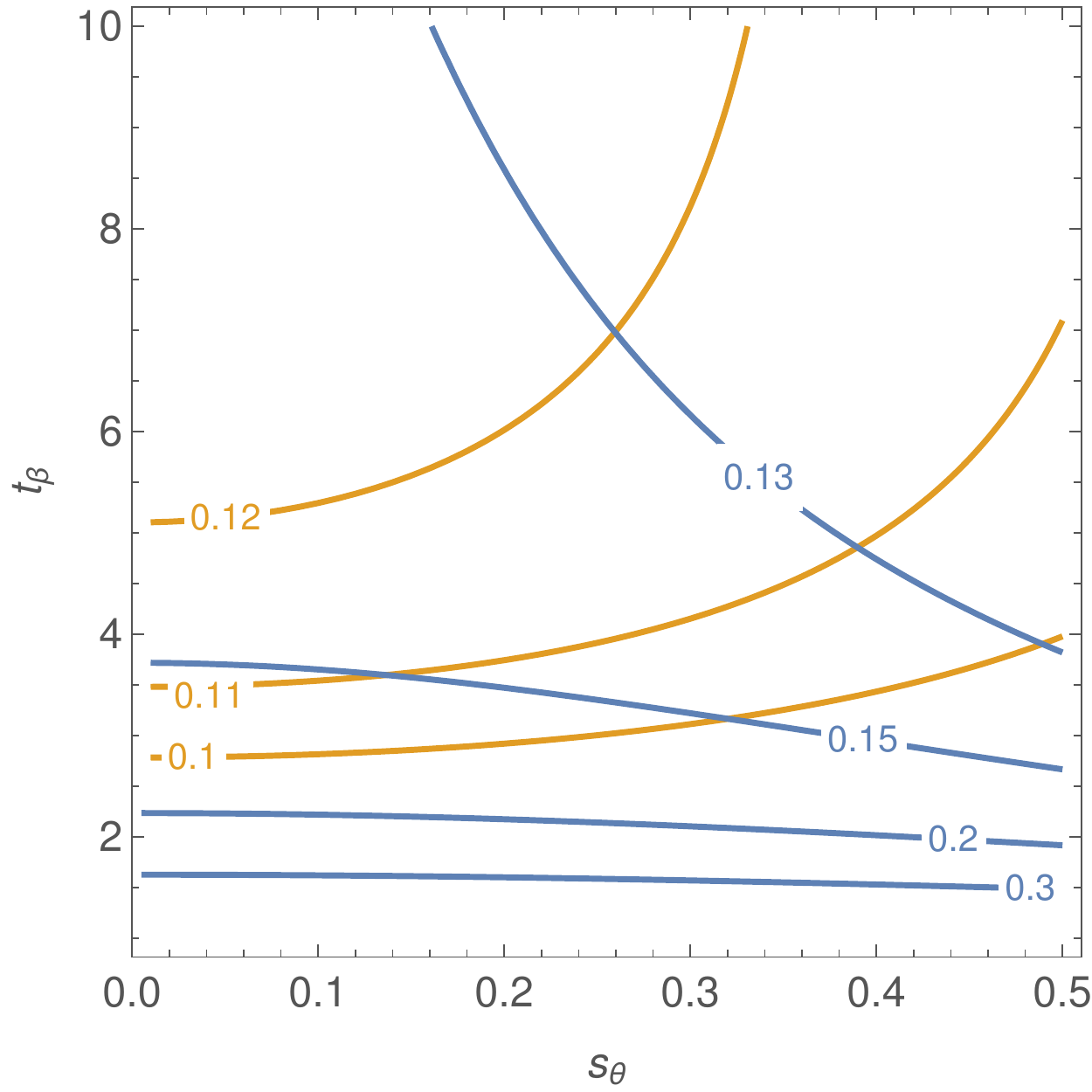}
	    \includegraphics[width=0.45\textwidth]{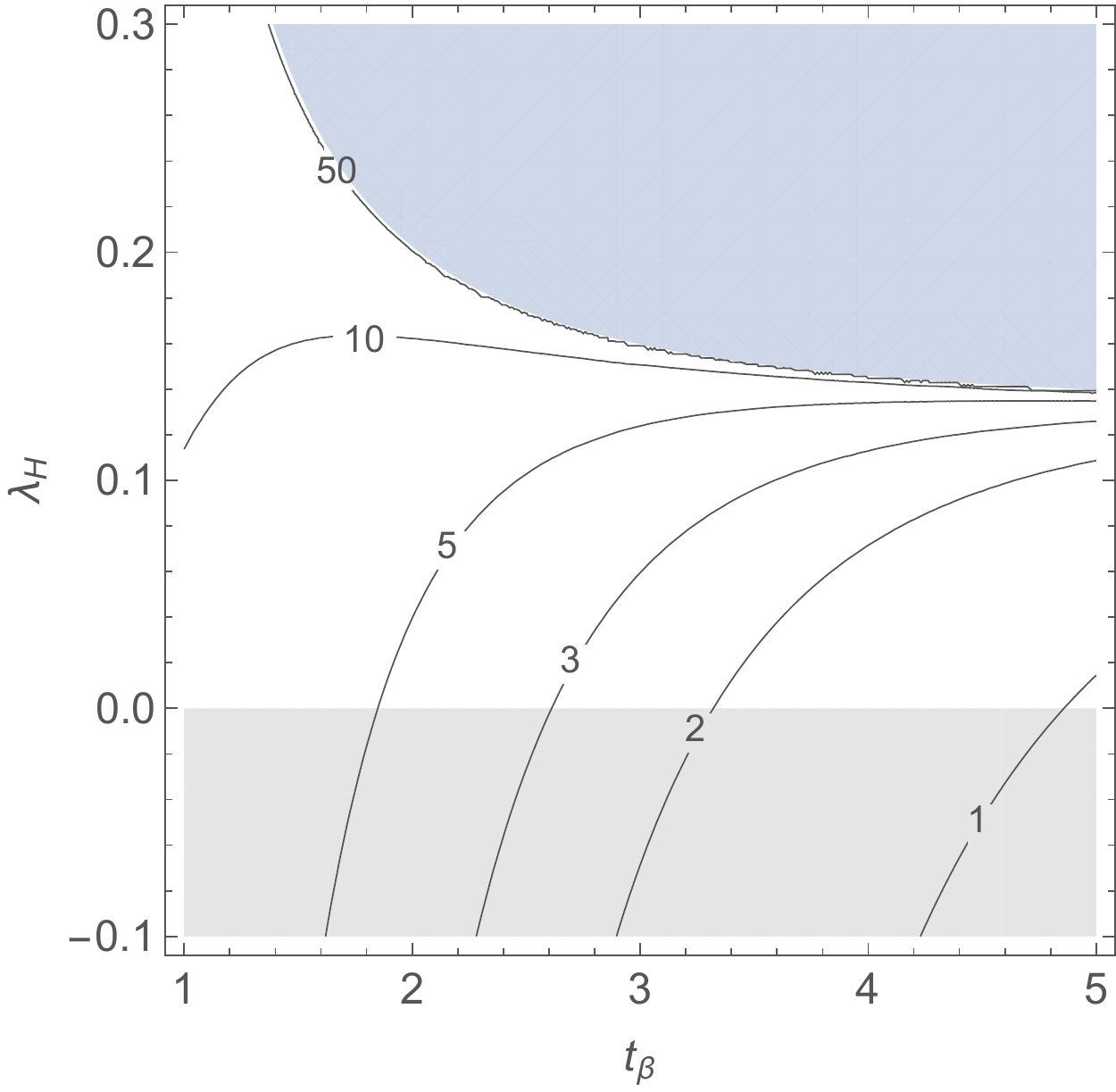}
	\end{center}
	\caption{
	{\bf Left panel:} Values of $\lambda_H$. Blue (yellow) contours represent the case ${m_H^2/f^2=0}$ (${m_H^2/f^2=1}$).
	{\bf Right panel:} The contours show the value of $\Lambda_{\mathrm{TC}}=4\pi f$ in TeV corresponding to a 1 TeV heavy Higgs mass.
	The region $\lambda_H<0$ is excluded by the 
	requirement of stability of the potential.  
	For  $\Lambda_{\mathrm{TC}}\rightarrow \infty$ the curves lie on top of the $\Lambda_{\mathrm{TC}}=50$ TeV contour 
	so the shaded area results in no viable solutions. More details in the text.  }
	\label{fig:rlambdaCt}
    \end{figure}

    Using the experimental value of the Higgs mass and fixing the mass parameter to $m_H^2=0$ ($m_H^2=f^2$),    
    we show on the left panel of Fig.~\ref{fig:rlambdaCt} with blue (yellow) contours the value of $\lambda_H$ in
    $(s_{\theta},t_{\beta})$ plane.
    The figure demonstrates that the value of $\lambda_H$ may only be increased slightly beyond its value in the 
    SM $\lambda_{\rm SM}\approx 0.129$ unless $t_{\beta}$ is very small.
    Since for large $t_{\beta}$, the light Higgs mass is a sum of two manifestly 
    positive contributions, cf.~\eq{eq:lighthiggs}, the scalar quartic coupling is bounded for a given value of $t_\beta$ to 
    $\lambda_H \lesssim \lambda_{\rm SM}/s_\beta^2$, and the bound is only reached in the  limit $m_H^2\to 0, s_\theta\to 0$.
    Furthermore, for $t_{\beta}\rightarrow \infty$, this bound reduces to the SM value.
    On the right panel of Fig.~\ref{fig:rlambdaCt}, we show the contour lines in the $(t_{\beta}, \lambda_H)$ plane for different 
    values of $\Lambda_{\mathrm{TC}}=4\pi f$ in TeV having fixed the Higgs mass, and the mass of the heavier eigenstate 
    to 1 TeV for concreteness. 
    In the lower shaded region, $\lambda_H<0$ and this is excluded due to vacuum 
    instability. The boundary of the upper blue shaded region shows the
    maximum allowed value of $\lambda_H$ for which the correct Higgs mass can still be achieved as a function of $t_{\beta}$. The  
    $\Lambda_{\mathrm{TC}}\rightarrow \infty$ contours lie very close to the $\Lambda_{\mathrm{TC}}=50$ TeV contour. This 
    boundary is almost independent of the fixed value of the heavy eigenstate mass, and the blue shaded region is thus inaccessible.
    
    The accessible values of $\lambda_H$ shown in Fig.~\ref{fig:rlambdaCt},
    together with the enhanced top Yukawa in Eq.~(\ref{eq:topyukawa}), make it
    clear that it is important to consider the  vacuum stability constraints on the model.

    \subsection{RG analysis and vacuum stability}
    \label{sec:RG}

    The running of $\lambda_H$ is driven dominantly by the top loop, as in the SM. However, compared to the SM, the top Yukawa coupling 
    of the elementary interaction eigenstate, $\sigma_h$, is enhanced by $y_t=y_t^{\mathrm{SM}}/s_{\beta}$. Therefore $\lambda_H$ will 
    run negative faster than in the SM for the same weak scale initial value, and the vacuum of the model 
    will, thus, not be stable in all of parameter space. 
    The problem is similar to that in bTC~\cite{Carone:2012cd}  and in type-I two-Higgs-doublet models (2HDM).

    In Appendix~\ref{app:RG-D}, we tabulate the 
    one-loop beta functions for the bTC/pCH framework  with $N_{\mathrm{F}}$ new $\SUL$-doublet fermionic fields, 
    $Q_{\mathrm{L}}=(U_{\mathrm{L}},D_{\mathrm{L}})$, transforming under the 
    representation $R_{\mathrm{F}}$ of the new strong gauge group. These fields will couple to 
    the elementary Higgs doublet, $H$, through Yukawa-like interactions with couplings $y_U$ and $y_D$. 

    In the numerical analysis 
    we assume that the top quark Yukawa and the EW gauge couplings run as in the SM 
    below the condensation scale $\Lambda_{\mathrm{TC}}$, and we neglect the threshold effects on the quartic coupling running.
    We fix the initial values of the SM gauge couplings at $\mu_Z=m_Z$, and 
    the value of the SM
    top Yukawa from $y_t^{\rm SM}(m_t)$ as in~\cite{Degrassi:2012ry}.
    Furthermore, at the condensation scale, $\Lambda_{\mathrm{TC}}$, 
    we fix the TC gauge coupling to the estimate of the  critical value~\cite{Marciano:1980zf},
    $\alpha_{\mathrm{TC}}^{\mathrm{c}}= \frac{\kappa}{C_2(R_F)}$,
    where $\kappa\sim\mathcal{O}(1)$, and we assume identical Yukawa couplings of the $U,D$
    technifermions $y_U(\Lambda_{\mathrm{TC}})=y_D(\Lambda_{\mathrm{TC}})\equiv y_Q^0$. 
    However, we note that since  $y_{UD}\ll 1$ throughout parameter space consistent with the vacuum and Higgs mass conditions,
    the contribution of the new fermions to the running of the quartic coupling is negligible.

    \begin{figure}
	\begin{center}
	    \includegraphics[width=0.45\textwidth]{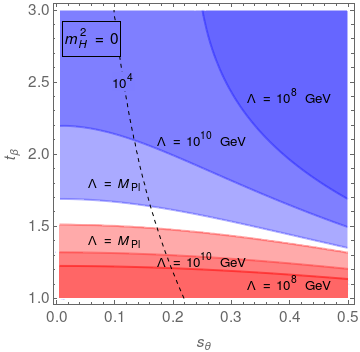}\
	    \includegraphics[width=0.45\textwidth]{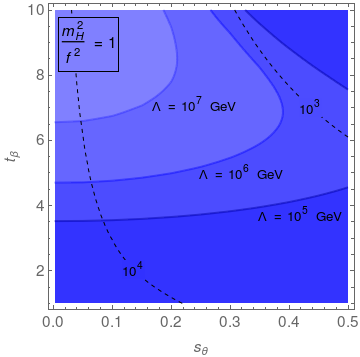}
	\end{center}
	\caption{The blue (red) regions illustrate the vacuum instability (non-perturbativity) scale. 
	The dashed lines show the values of $\Lambda_{\mathrm{TC}}=4\pi f$ in GeV. {\bf Left panel:} $m_H^2=0$. On the boundaries of the regions, the
	vacuum is stable (couplings pertubative) up to $\Lambda=M_{\mathrm{Pl}}=1.22\cdot 10^{19}~\mathrm{GeV},\ 10^{10}~\mathrm{GeV},\ 10^8~\mathrm{GeV}$. 
	{\bf Right panel:} $m_H^2/f^2=1$. On the boundaries of the regions, the	vacuum is stable up to $\Lambda=10^7~\mathrm{GeV},\ 10^6~\mathrm{GeV},\ 
	10^5~\mathrm{GeV}$.}
	\label{fig:window1}
    \end{figure}

    \begin{figure}
	\begin{center}
	    \includegraphics[width=0.7\textwidth]{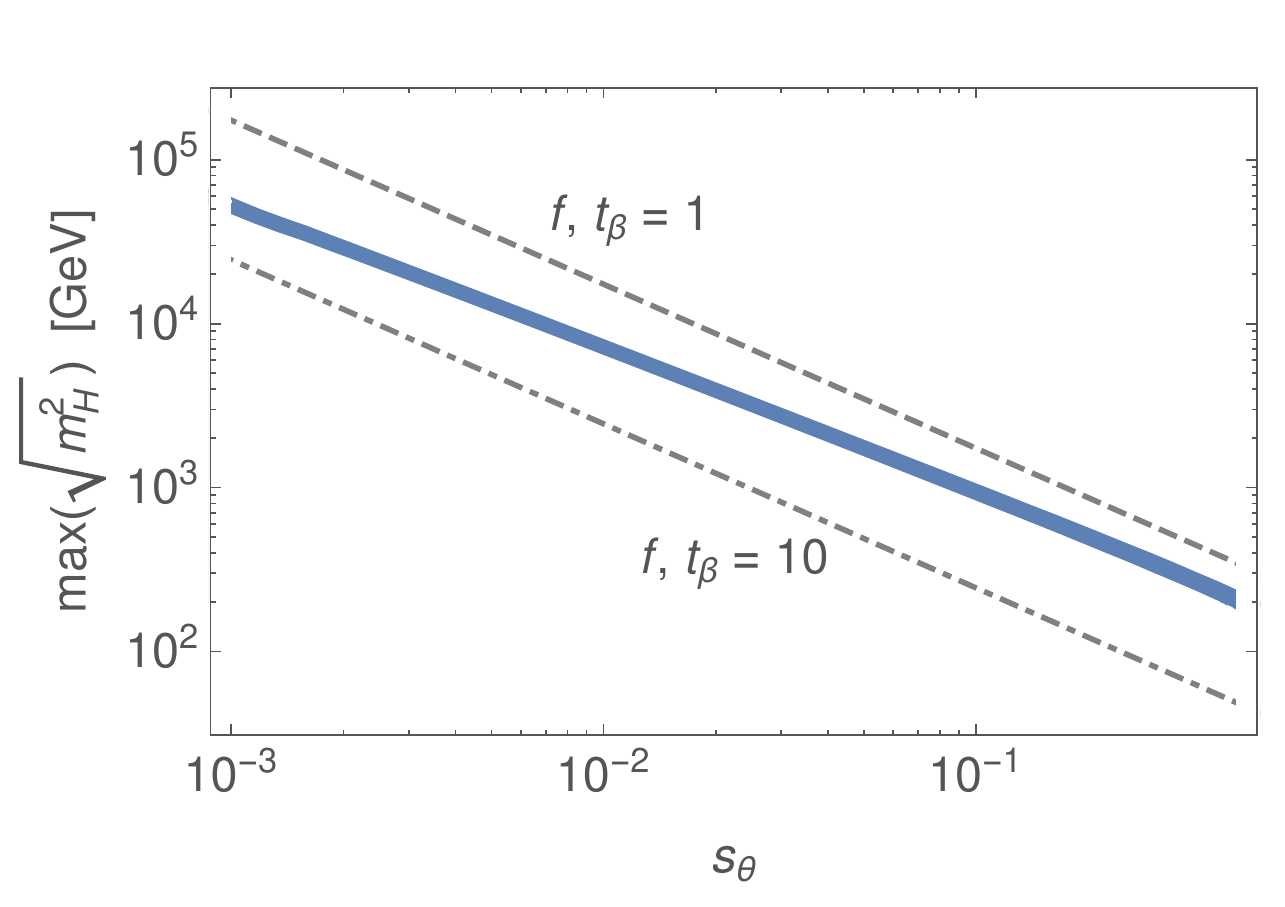}
	\end{center}
	\caption{ The blue band shows the maximum 
	allowed value of $\sqrt{m_H^2}$ as a function of $s_{\theta}$ if stability at least up 
	to the compositeness scale, 
	$\Lambda_{\mathrm{TC}}=4\pi f$, is required and  $t_{\beta}$ is varied within the range $1\dots 10$. The dashed 
	(dot-dashed) line shows $f$ as a function of 
	$s_{\theta}$ for constant value of $t_{\beta}=1$ ($t_{\beta}=10$). 
	}
	\label{fig:maxnat}
    \end{figure}

    In Fig. \ref{fig:window1}, we show the vacuum instability and perturbativity scales, $\Lambda_{\mathrm{stab}}, \Lambda_{\mathrm{pert}}$, resp.,
    defined through $\lambda_H(\Lambda_{\mathrm{stab}})=0$,  
    $\lambda_H(\Lambda_{\mathrm{pert}})= 4\pi$, in the $s_\theta, t_\beta$ plane for 
    two different values of $m_H^2/f^2=0$ and $m_H^2/f^2=1$. 
    Small values  $m_H^2/f^2 \ll 1$ correspond to large values of $\lambda_H$ as is evident from the 
    vacuum equations
    and yield a larger vacuum instability scale. 
    As seen in the left panel of Fig.~\ref{fig:window1}, there is a narrow window in parameter space with vacuum 
    stability and perturbativity up to the 
    Planck scale for relatively small $t_{\beta}$. For even smaller values of $t_{\beta}$, the quartic becomes non-perturbative, 
    whereas for larger values of $t_{\beta}$
    the initial value of $\lambda_H$ decreases, while the top coupling is still enhanced leading to vacuum instability 
    at lower scales. 
    Increasing $m_H^2/f^2$ lowers the viable values of $\lambda_H$, and this window is 
    lost as seen in the right panel of Fig.~\ref{fig:window1}. 
    
    The same increase in scale 
    happens when $s_\theta$ is reduced with $m_H^2$ and $t_{\beta}$ fixed. This corresponds to decreasing of $v/f$ and thus from 
    the vacuum conditions and the Higgs mass condition increasing $\lambda_H$.

    Overall, demanding a stable vacuum to high scales requires tuning  $m_H^2/f^2$ small. This implies 
    at least a modest hierarchy 
    between the scalar mass parameter and the compositeness scale.
    On the other hand, increasing the (absolute value of the) scalar mass parameter, $m_H^2$, with respect to the SM value requires 
    tuning $s_\theta\ll 1$ small.
   This must arise from balancing two sources of unrelated physics. 
    In Fig.~\ref{fig:maxnat}, we show the maximum allowed values of $m_H^2$ 
    as a function of $s_\theta$ demanding vacuum stability at least up to the compositeness scale, $\Lambda_{\rm TC}=4\pi f$. 
    This constrains the maximum possible $m_H^2$ regardless of the UV completion above the compositeness scale. 
    We emphasize that while tuning $s_{\theta}\ll 1$ allows for larger values of $m_H^2$, it does not increase 
    the maximum possible value of $m_H^2/f^2$. For example for $s_{\theta}=0.1$, and $t_{\beta}=1$ ($t_{\beta}=10$) we 
    find $\max\left( \frac{m_H^2}{f^2}\right)\approx 0.3$ (14.5), 
    whereas for $s_{\theta}=10^{-3}$, the corresponding value is 0.1 (5.7).

    \subsection{Additional source for the top-quark mass}
   \label{subsec:Add}

    As seen in the previous RG analysis, in the simplest partially composite scenarios, the top-Yukawa coupling is enhanced 
    with respect to the SM driving    
    the running quartic coupling to negative values before the Planck scale in most of parameter space. As seen in 
    Fig.~\ref{fig:window1}, there is a small region where the weak scale value of the quartic itself is sufficiently 
    enhanced with respect to the SM Higgs value, to counter balance the enhanced top Yukawa. A way to reduce the constraints 
    from vacuum stability is to assume the top quark mass is not
    entirely due to the elementary scalar vev, but acquires a contribution from some still unspecified dynamics. This implies 
    a reduction of the top Yukawa coupling of $\sigma_h$ in Eq.~(\ref{eq:topyukawa}).   

    To this end, we start by adding a four-fermion operator between the top and the techniquarks
    \begin{equation}
	\label{eq:1f}
	\begin{split}
	    \mathcal{L}_{\mathrm{4f}}\sim&-\frac{Y_tY_U}{\Lambda_t^2}(\bar{q}_{\mathrm{L}}t_{\mathrm{R}})^{\dagger}_{\alpha}
		(Q^{T}P_{\alpha}Q)+\ \mathrm{h.c.}
	\end{split}
    \end{equation} 
    We discuss the possible sources of the above four-fermion operator in the end of the section. 

    Upon the condensation of the techniquarks, this yields a contribution to the top mass, i.e.
    \begin{equation}
	\label{eq:L4f}
	    \mathcal{L}_{\mathrm{4f}}\sim-4\pi f^3 Z_2\frac{Y_tY_U}{\Lambda_t^2}\Tr[P_1\Sigma]\bar{t}t=-y_t^{\prime}f s_{\theta} \ \bar{t}t+\dots,
    \end{equation} 
    where 
    \begin{equation}
	\label{eq:}
	y_t^{\prime}\equiv\frac{4\pi f^2 Y_tY_U Z_2}{\Lambda_t^2}.
    \end{equation}
    Furthermore, this gives a contribution to the effective potential via the top loop
    \begin{equation}
	\label{eq:}
	    V_{\mathrm{top}}=-C_t y_t^{\prime\,2}f^4\left|\Tr[P_{\alpha}\Sigma]\right|^2=-C_ty_t^{\prime\,2}f^4 s_{\theta}^2+\dots
    \end{equation}
    Ignoring the subdominant EW-gauge contributions, the effective potential  becomes 
    \begin{equation}
	\label{eq:}
	V_{\mathrm{eff},\mathrm{top}}=V_{\mathrm{eff}}+ V_{\mathrm{top}}
    \end{equation}
    where $V_{\mathrm{eff}}$ is given in Eq.~(\ref{eq:effpot}). 
    Defining $\xi_t$ as the fraction of the top mass originated from this new contribution,
    \begin{equation}
	\label{eq:}
	y_t^{\prime}f s_{\theta} \equiv \xi_t m_t,
    \end{equation}
    the vacuum condition in $\theta$ is changed to 
    \begin{equation}
	\label{eq:minCond}
	\begin{split}
	    0=\left.\frac{\partial V_{\mathrm{eff},\mathrm{top}}}{\partial \theta}\right|_{\mathrm{vac}}=
	    \left.\frac{\partial V_{\mathrm{eff}}}{\partial \theta}\right|_{\mathrm{vac}}
	    -\frac{2f^2 C_t\xi_t^2m_t^2}{t_{\theta}}.
       \end{split}
    \end{equation}
    The neutral mass matrix becomes
	\begin{equation}
	    \begin{split}
	    \label{eq:Mpi2t}
	    M_{h,\mathrm{top}}^2=m_{\lambda}^2\left(\begin{array}{cc}
		    1+\delta & -c_{\theta}t_{\beta}\\
		     -c_{\theta}t_{\beta} &
		     t_{\beta}^2+2C_t\xi_t^2m_t^2/m_{\lambda}^2
		\end{array}\right).
	    \end{split}
	\end{equation}

    The effect of the additional contribution to the top mass has only a tiny effect on the vacuum alignment and the 
    scalar mass eigenvalues in most of the parameter space. Therefore, 
    for subdominant contribution to the top mass,
    the main effect is just a small reduction of the top-Yukawa coupling,  with Eq.~(\ref{eq:topyukawa}) modified into 
  \begin{equation}
	\label{eq:topyukawa2}
	y_t=y_t^{\mathrm{SM}}(1-\xi_t)/s_{\beta}.
    \end{equation} 
    However, in the simplest 
    UV-completion scenarios of the four-fermion operator,
    the effective treatment is valid only if the contribution to the top mass is very 
    small.

\begin{figure}
	\begin{center}
	    \includegraphics[width=0.45\textwidth]{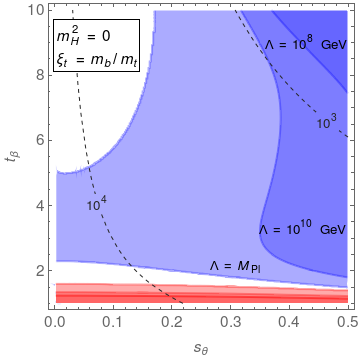}\ 
	    \includegraphics[width=0.45\textwidth]{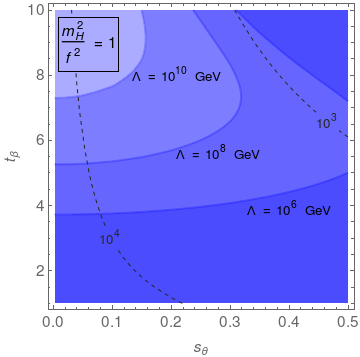}
	\end{center}
	\caption{Corresponding plots to Fig.~\ref{fig:window1} with $\xi_t=m_b/m_t$. On the unshaded area 
	    $\Lambda\geq M_{\mathrm{Pl}}$. The dashed lines show the values of $4\pi f$ in GeV.}
	\label{fig:windowTop1}
    \end{figure}

    A simple possibility to generate the four-fermion operator in Eq.~\eqref{eq:1f} is to add an additional heavy scalar doublet that couples 
    to the top quark and 
    to the new fermions and can be integrated out above the compositeness scale $\Lambda_{\mathrm{TC}}\sim 4 \pi f$ \cite{Simmons:1988fu}.
    Another possibility would be devising  extended-technicolor (ETC) interactions~\cite{Eichten:1979ah,Dimopoulos:1979es}. 
    As opposed to ordinary TC models,  ETC would only have to give a small contribution to the top mass. 
    This would allow a higher ETC scale and thus alleviate the constraints from  flavour-changing neutral currents. It would also potentially 
    simplify the ETC construction, having only to involve the third generation as in 
    e.g. Ref.~\cite{Cacciapaglia:2015yra}. A small reduction (at the per cent level), such as provided by a new isospin-symmetric sector 
    giving equal mass contributions
    to both top and bottom and providing all of the bottom mass  already soften the rapid decrease  of the quartic 
    coupling via the RG, thus enlarging the allowed range of parameter space  constrained by vacuum instability. The stability plots 
    for the value $\xi_t=m_b/m_t\simeq 0.025$ are shown in Fig.~\ref{fig:windowTop1}.
    The figures thus correspond to those in Fig.~\ref{fig:window1}, except that 
    part of the top mass now originates from the four-fermion operator above. 
    The situation is in fact similar to the SM case where 
    the uncertainty in
    the top mass could, in fact, bring the SM from meta-stable to stable.
 
    \subsection{Phenomenology}

    In this section we study constraints on the model arising from the modified Higgs couplings, from searches for new heavy scalars, and
    the new contributions to $B$-meson mixing.  
    
	\subsubsection{Higgs couplings}
	To study the couplings of the 125-GeV Higgs state, we first define the $\kappa_{f,i}$ and $\kappa_{V,i}$ coefficients, parameterising  
	the couplings of the mass eigenstate $h_i$, $i=1,2$ to SM fermions, $f$, and vector bosons, $V$, relative
	to the SM Higgs boson couplings
	 \begin{equation}
	    \label{eq:kappaDefi}
	    \kappa_{f,i}\equiv\frac{g_{h_i\bar{f}f}}{g_{h\bar{f}f}^{\mathrm{SM}}},	
	    \quad \kappa_{V,i}\equiv\frac{g_{h_iVV}}{g_{hVV}^{\mathrm{SM}}} ,\quad  i=1,2 . 
	\end{equation}
    
	The $h_1$ coefficients and their $ s_\theta^2,t_\beta^{-2} \ll 1 $ expansions to second order are given by 
	\begin{equation}
	\label{Eq:couplingsh1}
	    \begin{split}
		\kappa_{V,1} &= c_\alpha s_\beta - s_\alpha c_\beta c_\theta \ \to \  1- \frac{1}{2}s_\theta^2 t_\beta^{-2},  \\   
		\kappa_{f,1} &=  c_\alpha/ s_\beta  \ \to \ 1 + \frac{1}{2}s_\theta^2 t_\beta^{-2} - t_\beta^{-4} \delta \ .
	    \end{split}
	\end{equation}
	 The $h_2$ coefficients and their expansions to first order are given by
	 \begin{equation}
	 \label{Eq:couplingsh2}
	    \begin{split}
		\kappa_{V,2} &=s_\alpha s_\beta + c_\alpha c_\beta c_\theta \ \to \ \delta t_\beta^{-3}, \\  
		 \quad \kappa_{f,2} &= s_\alpha / s_\beta  \ \to \ t_\beta^{-1} 
		+ \delta  t_\beta^{-3} - \frac{1}{2} t_\beta^{-1} s_\theta^{2}.
	    \end{split}
	  \end{equation}
	The exact expressions above coincide with those derived in Ref.~\cite{Galloway:2016fuo} and with those in a type I 
	2HDM except 
	for the appearance of $c_\theta$.  
	
\begin{figure}
	    \begin{center}
		\includegraphics[width=0.45\textwidth]{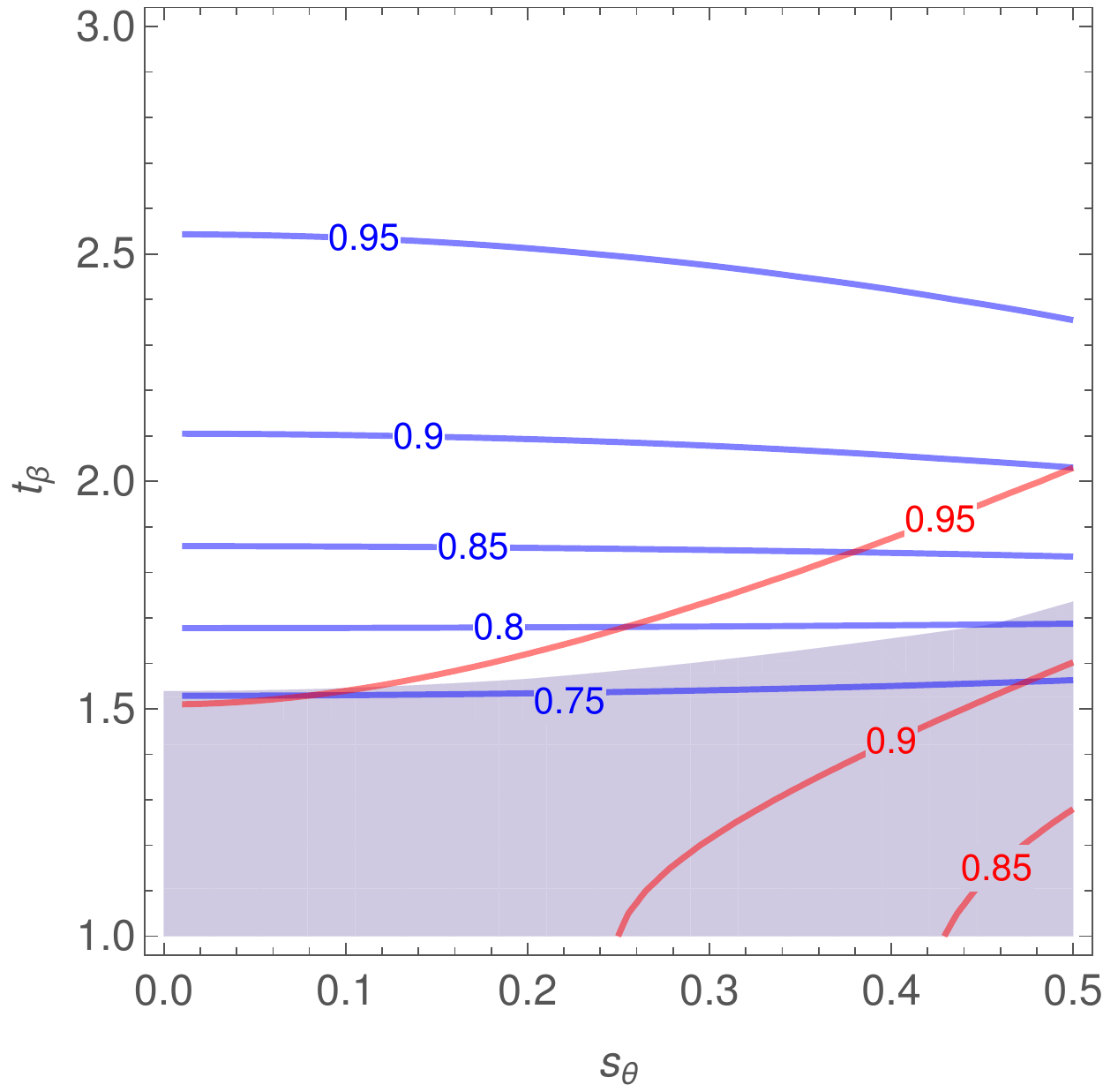}\
		\includegraphics[width=0.45\textwidth]{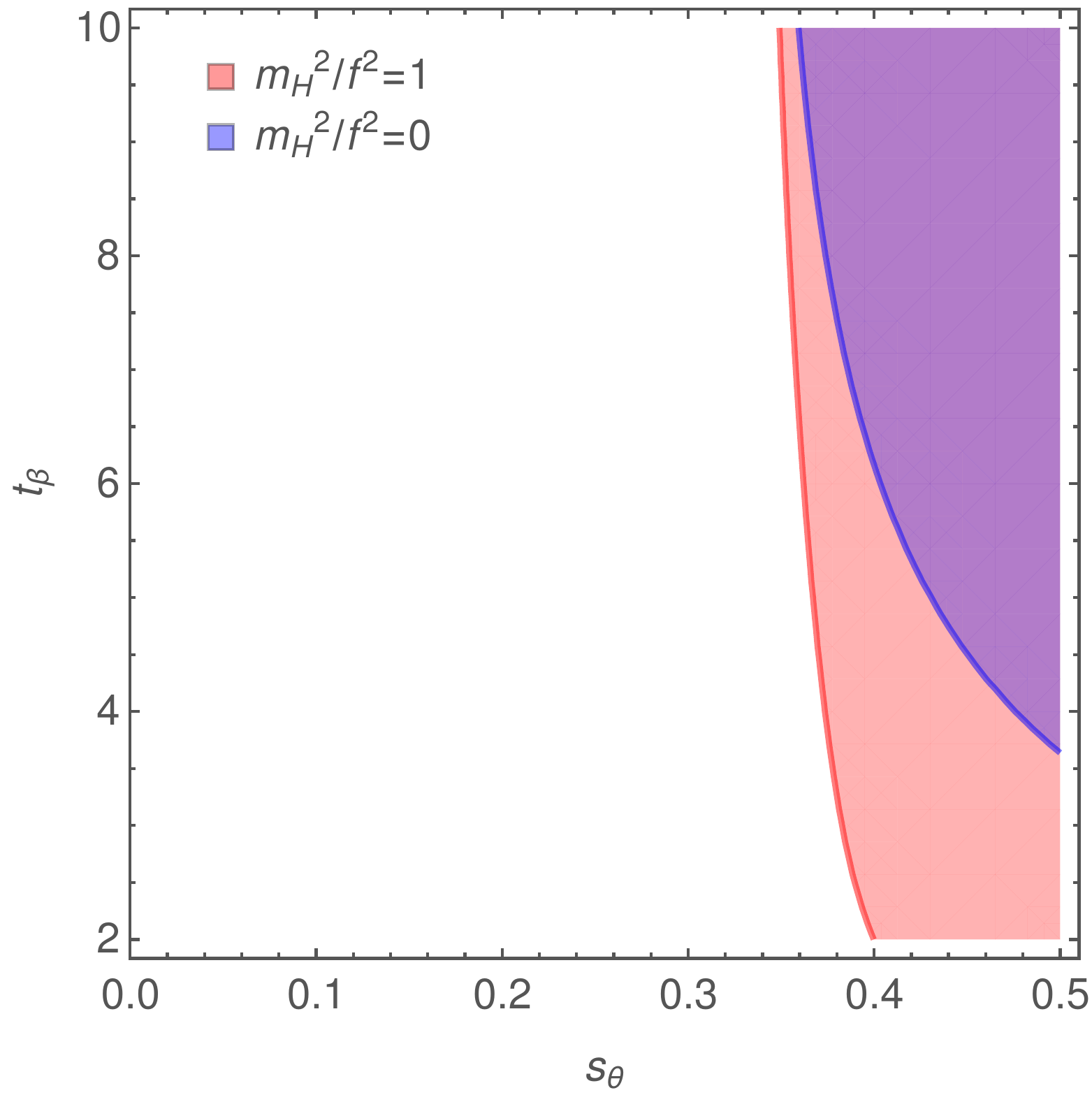}
	    \end{center}
	    \caption{{\bf Left panel:} The blue (red) curves show $|\kappa_t|$ ($|\kappa_V|$) for $m_H^2=0$ case. The shaded region shows 
		the $2\sigma$ exclusion from the two-parameter $\kappa_f,\kappa_V$ fit~\cite{Khachatryan:2016vau}. 
		{\bf Right panel:} Region in which $\kappa_b$ is within $2\sigma$ of the measured value $|\kappa_b|=0.57\pm 0.16$, 
		improved w.r.t. the SM prediction. }
	    \label{fig:kappas}
	\end{figure}

	Since the 125-GeV Higgs state is identified with $h_2$ instead of $h_1$ in the small part of parameter space with 
	$1+\delta > t_\beta^2$ as discussed below Eq.~\eqref{eq:lighthiggs}, we use $\kappa_{f}$ and $\kappa_V$ to denote the coupling 
	coefficients of the 125-GeV Higgs state in all of parameter space.
	We show the values of these coefficients in the left panel of 
	Fig.~\ref{fig:kappas} for $m_H^2=0$ ($\delta=2$) together with the region disfavoured at the $2\sigma$ level by LHC measurements 
	of the Higgs couplings from~\cite{Khachatryan:2016vau}. 
	In the parameter space shown, the deviation of $\kappa_f$ from unity is nearly independent of $s_\theta$ as follows from 
	Eqs.~(\ref{Eq:couplingsh1}) and~(\ref{Eq:couplingsh2}) when $\delta$ is sufficiently large. 
	In the very small $t_\beta$ region, the expansions are no longer applicable.

	The figure demonstrates how most of the considered parameter 
	space is still viable, but interestingly the classically scale-invariant region in Fig.~\ref{fig:window1} with vacuum stability up
	to the Planck 
	scale is just above what is currently probed by LHC Higgs coupling measurements. 
	For larger values of $m_H^2$ (smaller $\delta$) the deviations from the SM value are even smaller.

	In the case where we allow for a small part, $\xi_t$, of the top mass coming from a new sector encoded in the four-fermion interaction in 
	Eq.~\eqref{eq:1f}, the coupling of the top quark to Higgs is only slightly further modified:    
	\begin{equation}
	    \kappa_t = -\xi_t \frac{s_\alpha c_\theta}{c_\beta}+(1-\xi_t)\frac{c_\alpha}{s_\beta}\,.
	\end{equation} 
	A motivated possibility would be a new isospin-symmetric sector providing equal mass contributions to the third generation 
	and providing all of the bottom mass. In this case $\xi_t= m_b/m_t$, and this would result in a significant modification in  
	the couplings of the partially composite Higgs to bottom quarks:
	\begin{equation}
	    \kappa_b=-\frac{s_\alpha c_\theta}{c_\beta}\,.
	\end{equation}
	Interestingly, the measured value of this coupling by ATLAS and CMS combined is $|\kappa_b|=0.57\pm 0.16$~\cite{Khachatryan:2016vau} and 
	indicates 
	a reduction of approximately 2.7$\sigma$ with respect to the SM prediction according to the estimate which allows beyond-the-SM 
	contribution in the loops. 
	In the right panel of \fig{fig:kappas}, we show the favoured region in which $|\kappa_b|$ is within the $2\sigma$ region of 
	the measured value, for $m_H^2=f^2$, 
	and $m_H^2=0$. 	This reduction favours large values of $s_{\theta}$, which implies a tension with the requirement of vacuum 
	stability up to high scales. 

	\subsubsection{Scalar production}

\begin{figure}
	    \begin{center}
		\includegraphics[width=0.9\textwidth]{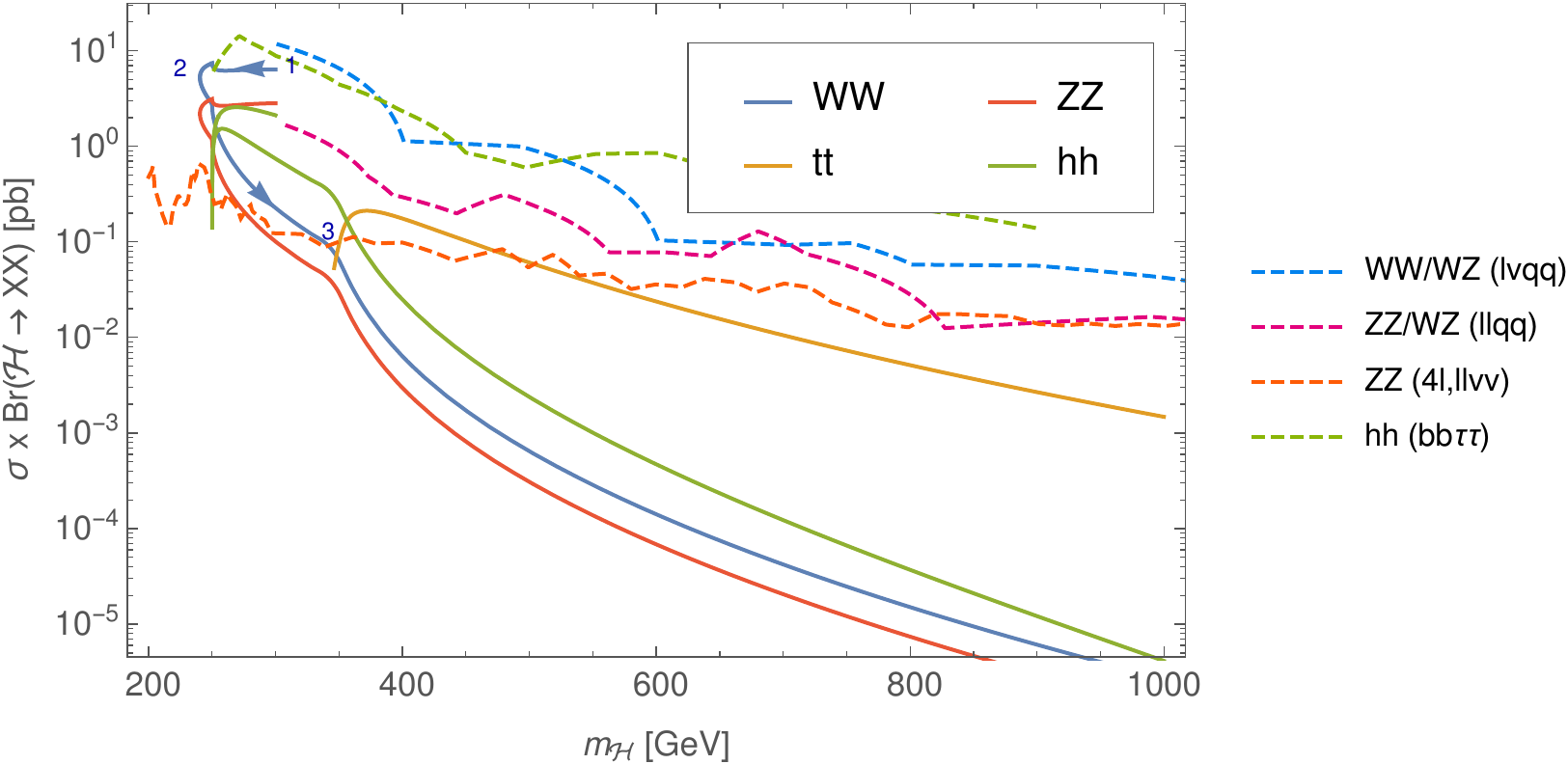}
	    \end{center}
	    \caption{$\sigma\times \mathrm{Br}({\cal H}\rightarrow XX)$ for the benchmark scenario with fixed $s_{\theta}=0.1$ 
		and $m_H^2=0$. The arrow shows the growing $t_{\beta}$, and the points 1, 2, and 3 represent $t_{\beta}=1,1.7,3.6$, resp. 
		The LHC limits are shown in dashed lines.}
	    \label{fig:h2prod}
	\end{figure}

	First, we study the production of the heavier CP-even eigenstate, which we denote by ${\cal H}$ as above, at the LHC. 
	Although the top coupling dominates the branching ratios parametrically, we find that the vector decay modes yield the strongest constraint. 
	We show the limits in Fig.~\ref{fig:h2prod} and find again that most of the parameter space of the model is viable. However, the LHC search 
	for ${\cal H}$ production in the $ZZ$ channel is in fact able to rule out the region stable up to the Planck scale for $m_H^2=0$. 

	The excluded cross sections at 95\% CL shown in the figure correspond to analyses using collisions at 13 TeV in the center-of-mass energy 
	to search for heavy resonances in $WW/WZ$ production in the decay channel $\ell \nu qq$~\cite{ATLAS:2017xvp} (blue), $ZZ$ production in the 
	decay channels
	$\ell^+\ell^-\ell^+\ell^-$ and $\ell^+\ell^-\nu\bar\nu$~\cite{ATLAS:2017spa} (orange), $ZZ$ and $ZW$ production in the 
	decay channels $\ell\ell qq$ and $\nu\nu qq$ \cite{Aaboud:2017itg} (red) and Higgs boson pairs decaying 
	into 
	$bb\tau\tau$~\cite{Sirunyan:2017djm} (green). 
	The decays of the light Higgs have been approximated by the SM prediction.
	The limits are shown in dashed lines.                        

	An interesting point to note is that the $t\bar{t}$ decay channel dominates above threshold, 
	cf. Eq.~\eqref{Eq:couplingsh2}. Resonant top-pair production is, 
	however difficult
	to compute precisely due to large interference effects 
	and QCD corrections~\cite{Bernreuther:2015fts,Hespel:2016qaf,BuarqueFranzosi:2017jrj}.  
	The potential of the LHC experiments to observe 
	this signal will rely on the ability to measure not only peak, but also other lineshape structures, such as peak-dip and pure dips. 
	This search is also one of the most important programs of the LHC, since the top quark may well be related to new physics and has the 
	least constrained interactions. 
	
	In the model an extra complication, which could in fact strengthen the 
	signal, is the presence of near degenerate 
	resonances: the three states, ${\cal H}$, $\pi^0$ and $\Pi_5$  have differences in mass of the order 
	of few GeV, as can be seen 
	in the left panel of \fig{fig:spectrum}, and thus the peak of the top pair production cross-section is expected to be 
	enhanced---although we note that the $\Pi_5$ decays into top quarks only via higher-scale physics~\cite{Arbey:2015exa}. 
	Top quark pair production in the CP-conserving Type-II 2HDM  with degenerate 
	masses $m_H=m_A$ and including interference effect has been considered by the ATLAS collaboration  at $\sqrt{s}=8$ TeV~\cite{Aaboud:2017hnm}. 
	In the limit of small $s_{\theta}$, the model here is well approximated by this case.
	The
	exclusion limit  at 95\% CL with $m_H=m_A=500$ GeV is $t_{\beta}=1.5$, and the exclusion limit decreases 
	to approximately  $t_{\beta}=0.9$ 
	for $m_H=m_A=600$ GeV. 
	At these mass values, the $t_\beta$ values in the pCH models considered here are still significantly higher (cf. Fig.~\ref{fig:spectrum}), 
	but we expect improved results from Run II. 
	For heavier masses, the analysis in Ref.~\cite{Aaboud:2017hnm} is not optimal, and the boosted regime must be considered. 
	In Ref.~\cite{BuarqueFranzosi:2017qlm} the LHC reach was studied for this kind of model through a reinterpretation of the top pair 
	differential cross section measurement in both the boosted and resolved regimes. Although the near degenerate case was not 
	considered, it seems to indicate that a 
	very low systematic uncertainty must be achieved.

	\subsubsection{$B^0-\overline{B^0}$  mixing}
\begin{figure}
	    \begin{center}
		\includegraphics[width=0.45\textwidth]{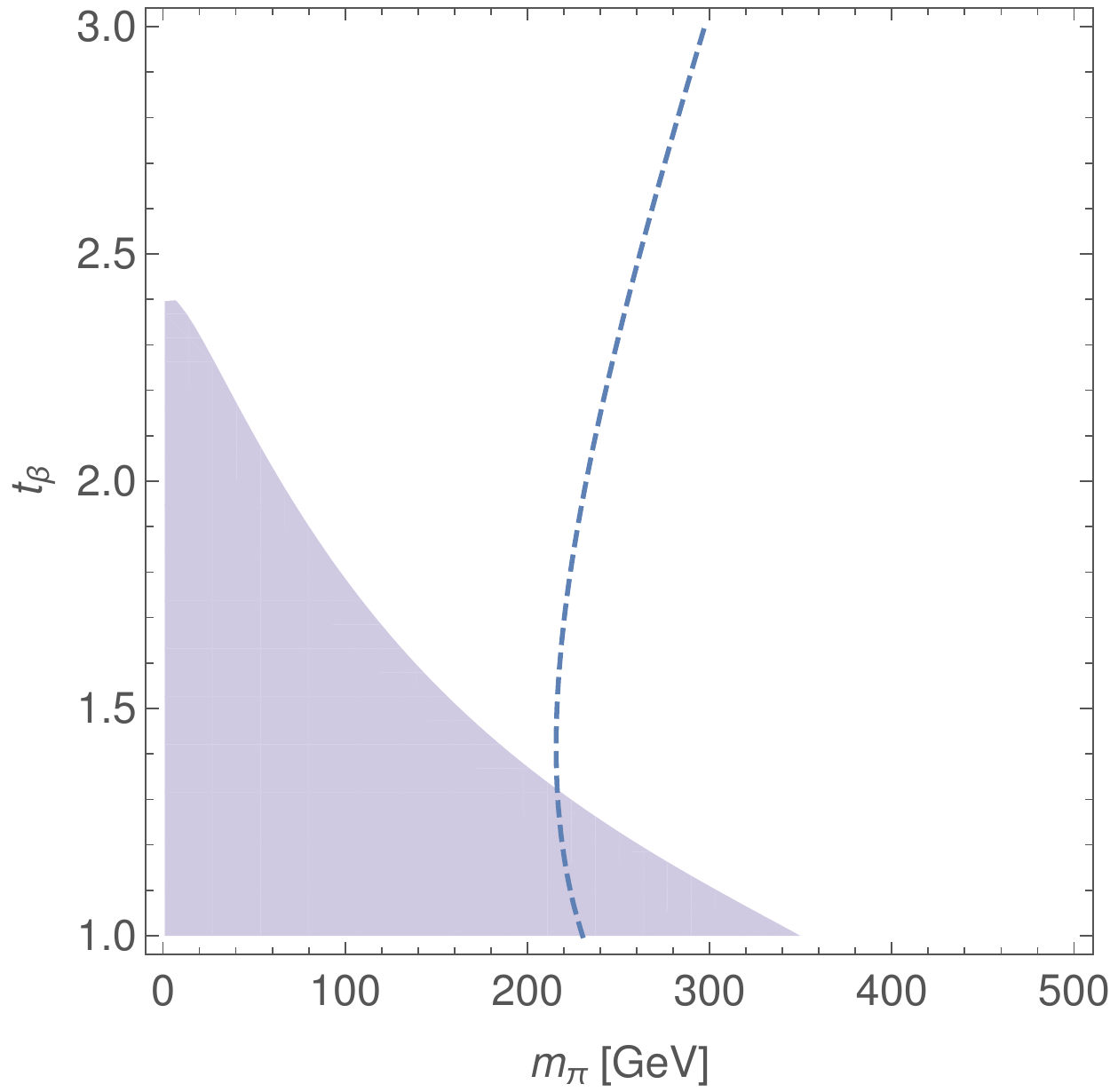}
	    \end{center}
	    \caption{Excluded region by $B^0-\overline{B^0}$ mixing. The dashed curve shows the value of $m_{\pi}$ calculated for the benchmark
	    scenario with $m_H^2=0$.}
	    \label{fig:B0exclusion}
	\end{figure}

	Models with two Higgs doublets, including bTC models~\cite{Simmons:1988fu,Carone:2012cd}, are in general constrained by measurements
	of $B^0-\overline{B^0}$ mixing. 
	 Following~\cite{Urban:1997gw}, we calculate the contribution from the heavy pions to $B^0-\overline{B^0}$ mixing taking into 
	account the next-to-leading order (NLO) QCD corrections. The mass splitting can be written as
	\begin{equation}
	    \label{eq:DeltamB}
	    \Delta m_B=\frac{G_{\mathrm{F}}^2}{6\pi^2}m_W^2|V_{td}V^*_{tb}|^2S(x_W,x_{\pi})\eta(x_W,x_{\pi})B_B f_B^2 m_B,
	\end{equation}
	where $x_{W,\pi}\equiv m_t^2/m_{W,\pi}^2$. The lengthy expressions for the functions $S$ and $\eta$ can be found 
	in Ref.~\cite{Urban:1997gw} and we use the current QCD lattice estimate of the decay constant 
	$f_B\sqrt{B_B}=216\pm 10\, \mathrm{MeV}$~\cite{Carrasco:2013zta}.
	Using the experimental result
	$\Delta m_B = (3.3321\pm 0.0013)\cdot10^{-13}~\mathrm{GeV}$~\cite{Patrignani:2016xqp}, we show the $2\sigma$ exclusion region in 
	the $(m_\pi, t_\beta)$ plane, along with the result for the pion mass 
	for $m_H^2=0$ in Fig.~\ref{fig:B0exclusion}.
	The figure shows that the region with low $t_{\beta}\lesssim 1.5$  is in tension with also the $B^0-\overline{B^0}$ results. 

\FloatBarrier

\section{Model II: RG analysis and phenomenological constraints}

We now consider the analysis of our Model II where the $\SU(2)$ doublet of elementary scalars is extended to a full $\SU(4)$ 
multiplet $\Phi$. We refer to Sec.~\ref{sec:modelII} for a short review of the model and the effective potential. 

    \subsection{Vacuum and spectrum}

    Following Ref.~\cite{Alanne:2017rrs}, we proceed to minimize the effective potential given in Eqs.~\eqref{eq:VeffSU4}--~\eqref{eq:VDeltam}. 
    This yields equations of constraint for the $\SU(4)$ invariant Yukawa coupling, $y_Q$, the vev of the singlet scalar, $v_S$,
    and the $\SU(4)$-breaking mass parameter, $\delta m^2$:
    \begin{equation}
	\label{eq:SU4BrVac}
	    y_Q=\frac{m^2_{\lambda} v}{8\sqrt{2}\pi Z_2 f^3 s_{\theta}},\quad
	    v_S=\frac{\widetilde{C}_g Z_2^2f^4s_{\theta}^2-v^2m_{\lambda}^2}{t_{\theta}vm_{\lambda}^2},\quad
	    \delta m^2=\frac{\widetilde{C}_g Z_2^2f^4s_{\theta}^2m_{\lambda}^2}{v^2m_{\lambda}^2-\widetilde{C}_g Z_2^2f^4s_{\theta}^2}\, ,
    \end{equation}
    where $v\equiv\langle\sigma_h\rangle$, $v_S\equiv \langle S_{\mathrm{R}}\rangle$, 
    $m^2_{\lambda}\equiv m_{\Phi}^2+\lambda_{\Phi}(v^2+v_S^2)$, and $\widetilde{C}_g\equiv C_g(3g^2+g^{\prime\,2})$.

    The three CP-even neutral states $\sigma_h, \Pi_4, S_{\mathrm{R}}$ mix, but with the $\delta m^2$ perturbation in the scalar 
    potential, we do not have simple analytical formulas for the mass mixings. We therefore solve numerically the rotation matrix, $R$, defined by 
    \begin{equation}
	\label{eq:rotationSU4}
	\left(\begin{array}{c}
	h_1\\
	h_2\\
	h_3
	\end{array}\right)\equiv
	R\left(\begin{array}{c}
	\sigma_h\\
	\Pi_4\\
	S_{\mathrm{R}}
	\end{array}\right).
    \end{equation}

    The lightest mass eigenstate, $h_1$, is identified with the
    observed Higgs boson with mass $m_{h_1}=125$~GeV. The masses of the physical heavy pion triplet, $\pi^{\pm,0}$, orthogonal to those 
    eaten by the $W$ and $Z$ bosons are given by
   
    \begin{equation}
	\label{eq:piMass}
    (m_{\pi}^{\pm,0})^2=\frac{8\sqrt{2}\pi Z_2\vw^2 y_Q}{t_{\beta}s_{\theta}^2}.   
    \end{equation}
    There are two additional CP-odd states in the spectrum: the mass eigenstates composed of $\Pi_5$ and $S_{\mathrm{I}}$.   
    The spectrum for fixed $s_{\theta}=0.1$ is shown in Fig.~\ref{fig:spectrum}.

    As shown in~\cite{Alanne:2017rrs}, it is also in this model possible to achieve the EWSB driven purely by the strong dynamics 
    with $m_\Phi^2>0,\delta m^2>0$. 
    In particular, in the viable parameter space able to produce the correct Higgs mass, the quartic coupling is generically enhanced 
    with respect to the SM value. Furthermore, $\delta m^2\sim m_\Phi^2\sim f^2$, and thus for $s_{\theta} \lesssim 0.1$
    the scalar mass parameters reach above 1 TeV.

    \subsection{RG analysis}
    \label{sec:RGSU4}

    The values of the new Yukawa couplings, $y_Q$, for
    successful vacuum solutions are very small, $y_Q\ll 1$ throughout the parameter space that we consider.  We therefore neglect their 
    contributions to the one-loop $\beta$-functions of the model given in App.~\ref{app:RG-SU4}. 
    
\begin{figure}
	\begin{center}
	    \includegraphics[width=0.44\textwidth]{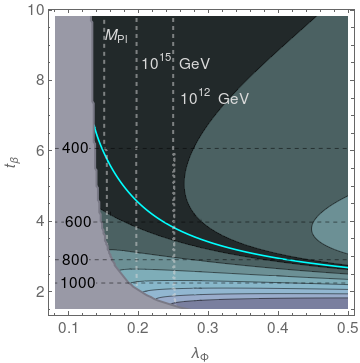}
	     \includegraphics[width=0.54\textwidth]{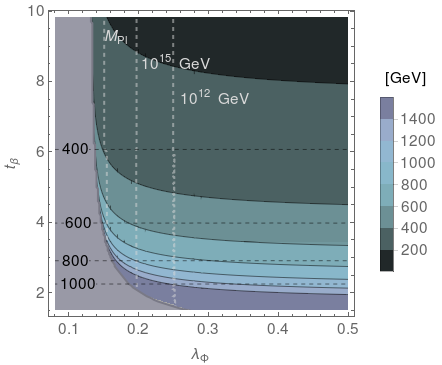}
	\end{center}
	\caption{The values of mass parameters $\sqrt{|m_{\Phi}^2|}$ ($\sqrt{\delta m^2}$) are shown in left (right) panel 
	when the Higgs mass is imposed for fixed $s_{\theta}=0.1$. Below the cyan line $m_{\Phi}^2>0$,  while 
	$\delta m^2>0$ throughout. On the gray region 
	on the left, no solution to the Higgs-mass constraint is found. See Ref.~\cite{Alanne:2017rrs} for details. 
	The dashed vertical lines show the parameter values where the maximum scale up to which the perturbativity of 
	the coupligns can be attained is  $10^{12}$~GeV, $10^{15}$~GeV or $M_{\mathrm{Pl}}$. 
	}
	\label{fig:SU4pert}
    \end{figure}

    The vacuum stability 
    is improved with respect to Model I due to the enhanced quartic coupling, $\lambda_\Phi$, in most of the parameter space. 
    However, the extended scalar sector now plays an important role because the SM-singlet elementary scalars, $S_{\mathrm{I},\mathrm{R}}$, 
    do not couple to SM fermions, and due to $y_Q\ll 1$ only very feebly to the techniquarks. Their quartic self-couplings may therefore develop
    Landau poles, and in fact the bounds on parameter space from perturbativity of these couplings are more relevant than that 
    coming from requiring vacuum stability. 
    We calculate the scale, $\Lambda$, at which at least one of the couplings exceeds $4\pi$, and plot the contours of 
    $\Lambda=M_{\mathrm{Pl}},\ 10^{15}\,\mathrm{GeV},\ 10^{12}\,\mathrm{GeV}$ as dashed vertical lines
    in Fig.~\ref{fig:SU4pert} along with the mass parameters $m_{\Phi}^2$ and $\delta m^2$. Below the cyan line on the left panel of
    Fig.~\ref{fig:SU4pert}, $m_{\Phi}^2>0$, while $\delta m^2>0$ throughout, and the EWSB is completely induced by the composite sector.
    On the shaded gray region, the correct Higgs mass cannot be achieved. See Ref.~\cite{Alanne:2017rrs} for details.
    
    One possibility to tame the running of the quartic couplings of the EW-singlet scalars $S_{\mathrm{I},\mathrm{R}}$ is to couple these to 
    right-handed neutrinos  $\nu_R$  with $\lesssim\mathcal{O}(1)$ Yukawa couplings. This will drive the singlet quartics down, and 
    result in higher perturbativity scale, as shown in Ref.~\cite{Alanne:2014bra} in a generic framework with one singlet scalar 
    and singlet fermion. Interestingly, in this case one family of leptons $e_L, \nu_L,  e_R^*, \nu_R$ may be embedded into a full $\SU(4)$ 
    multiplet under the global symmetry analogously to the technifermions, as done in e.g. Ref.~\cite{Foadi:2007ue}.
    The vev in the singlet 
    direction would in this case give dynamical Majorana masses for the  $\nu_R$'s, while the vev in the doublet direction would yield a Dirac 
    mass term. We leave the specifics of this construction for future work.

    \subsection{Phenomenology}
    \label{sec:colliderSU4}

    The constraints from $B^0-\overline{B^0}$  mixing are not relevant in the parameter space considered in Fig.~\ref{fig:SU4pert} 
    because the scalar masses are now heavy in the range where $t_\beta$ is small and vice versa. 
    The functions $S$ and $\eta$ in Eq.~(\ref{eq:DeltamB}) are therefore always small. 
    The $\kappa$-coefficients for this model, defined as in Eq.~\eqref{eq:kappaDefi} but now with $i=1,2,3$,
    can be written in terms of the rotation matrix 
    defined in Eq.~\eqref{eq:rotationSU4} as
    \begin{equation}
	\label{eq:kappaSU4}
	\kappa_{f,i}=\frac{R_{i1}}{s_{\beta}},\quad
	\kappa_{V,i}=R_{i1}s_{\beta}+R_{i2}c_{\beta}c_{\theta} ,\quad  i=1,2,3 . 
    \end{equation}
    We refer to Ref.~\cite{Alanne:2017rrs}  for an explicit illustration of the $\kappa_{f,1}$-coefficient for fixed 
    value of $s_{\theta}=0.1$. We conclude that for the most of the parameter space, the model is within the $2\sigma$
    region of two-parameter fit of $\kappa_{f,1}, \kappa_{V,1}$ on the combined CMS+ATLAS Run 1 data~\cite{Khachatryan:2016vau}.
    
\begin{figure}
	\begin{center}
	    \includegraphics[width=0.9\textwidth]{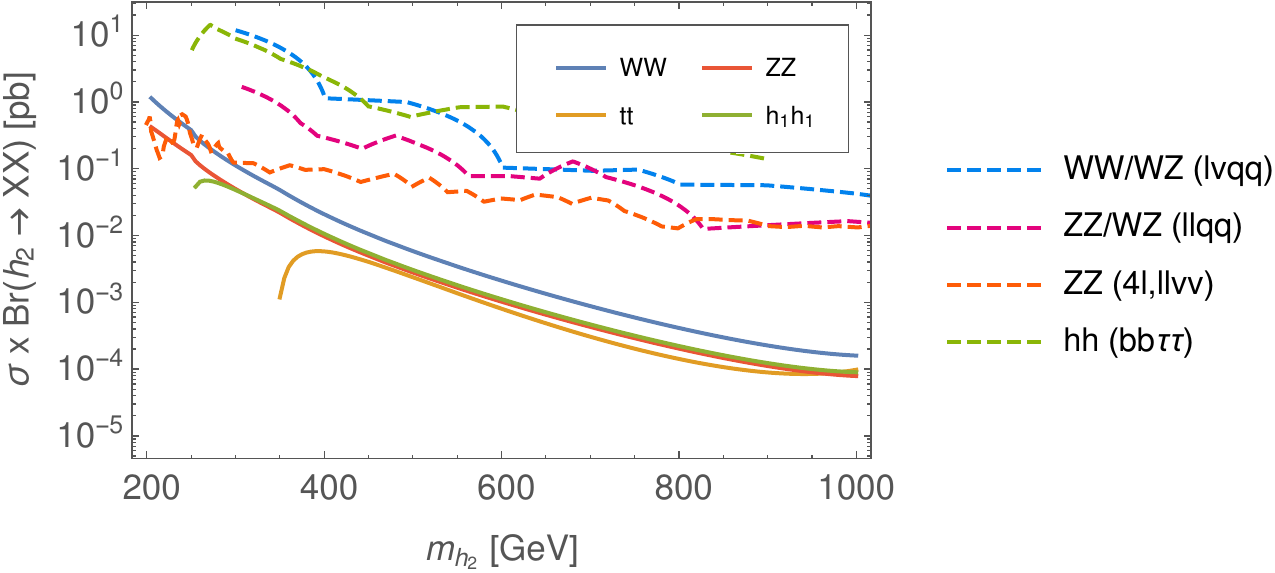}
	\end{center}
	\caption{$\sigma\times \mathrm{Br}(h_2\rightarrow XX)$ in different channels with corresponding collider limits. We have fixed $\lambda_{\Phi}=0.15$ 
	and $s_{\theta}=0.1$.}
	\label{fig:collimSU4}
    \end{figure}

    We present the collider limits relevant to the next to lightest neutral CP-even scalar, $h_2$, in Fig.~\ref{fig:collimSU4}. 
    The constraint from searches for heavy neutral scalars in the $ZZ$ channel is reduced as compared to the $\SU(2)$-doublet-scalar case due 
    to the additional mixing with the singlet $S_{\mathrm{R}}$. At face value, disregarding the fluctuations at small masses, the limit 
    is about 225 GeV. 
    In particular we note that the top coupling is significantly suppressed as compare to the Model I,  
    and even at large mass the $t\bar{t}$ production is subdominant. 

    \begin{figure}[b]
	\centering
	\begin{subfigure}{0.48\textwidth}
	\centering
	    \includegraphics[width=0.95\textwidth]{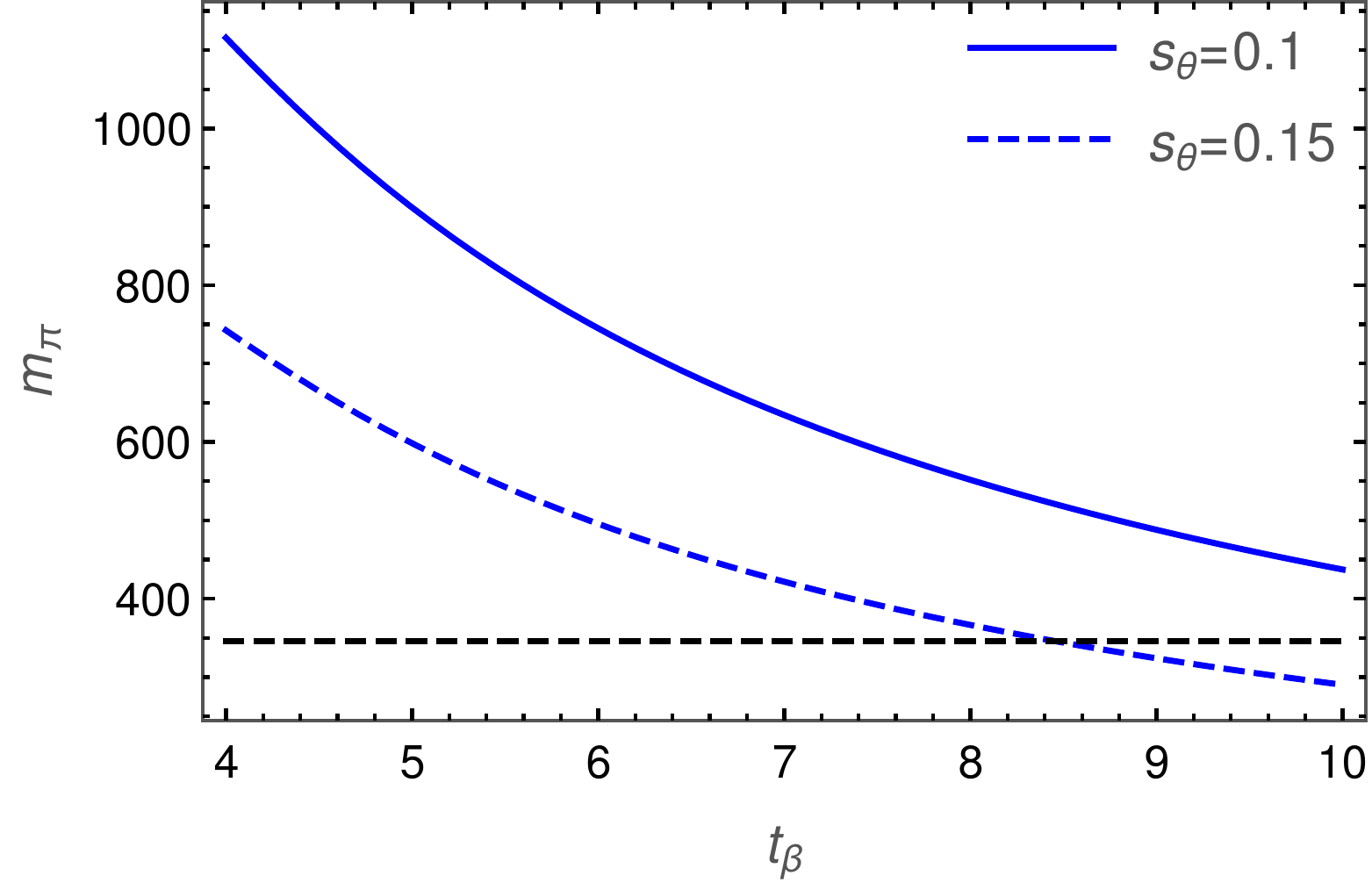} 
	 \caption{}
	\label{fig:h3pi3-mass}
	\end{subfigure}\
	\begin{subfigure}{0.48\textwidth}
	    \includegraphics[width=0.95\textwidth]{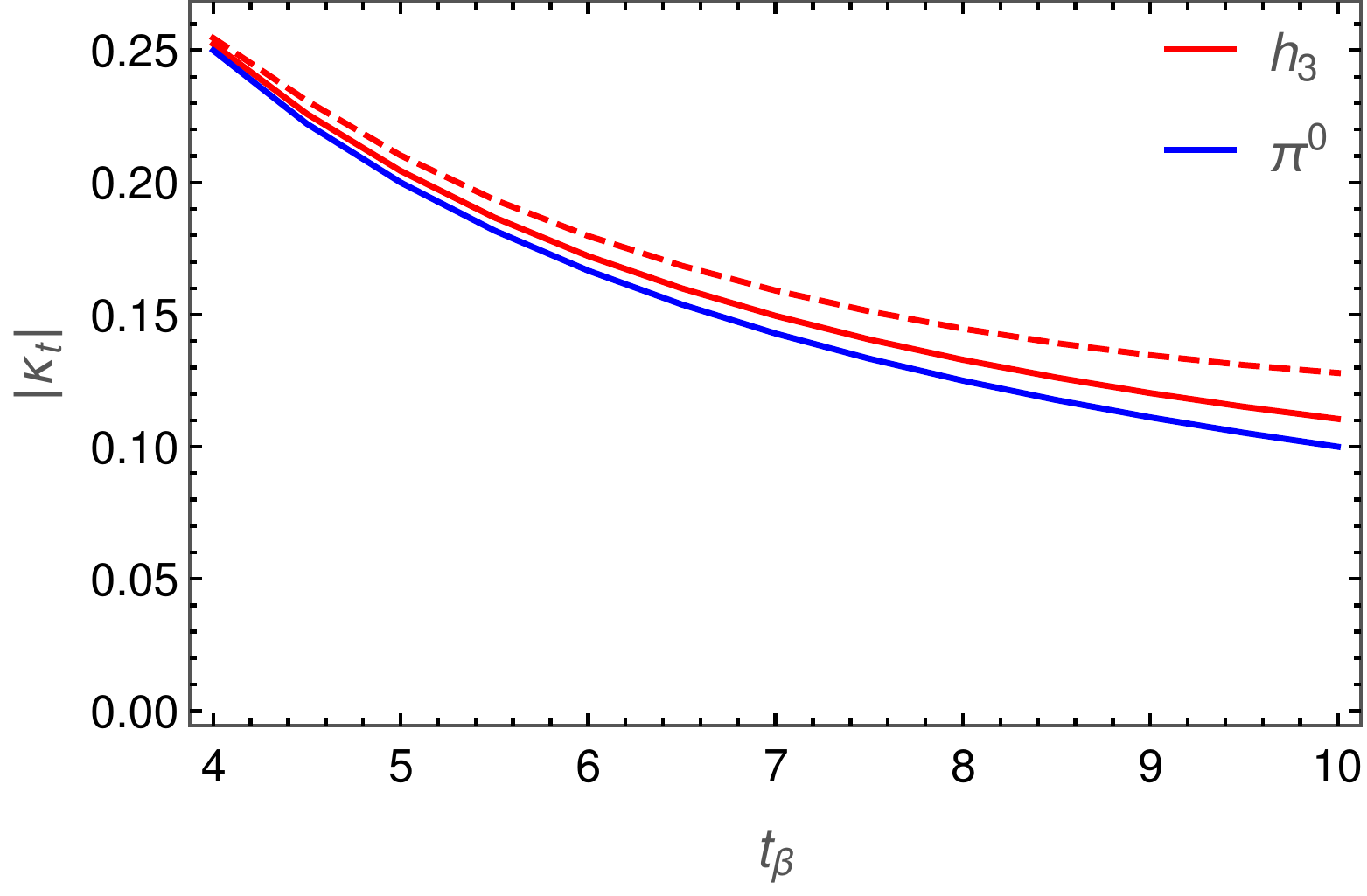}
	\centering
	 \caption{}
	\label{fig:h3pi3-kt}
	\end{subfigure}\\
	\begin{subfigure}{0.55\textwidth}
	\centering
	    \vspace{0.4cm}
	   \includegraphics[width=0.95\textwidth]{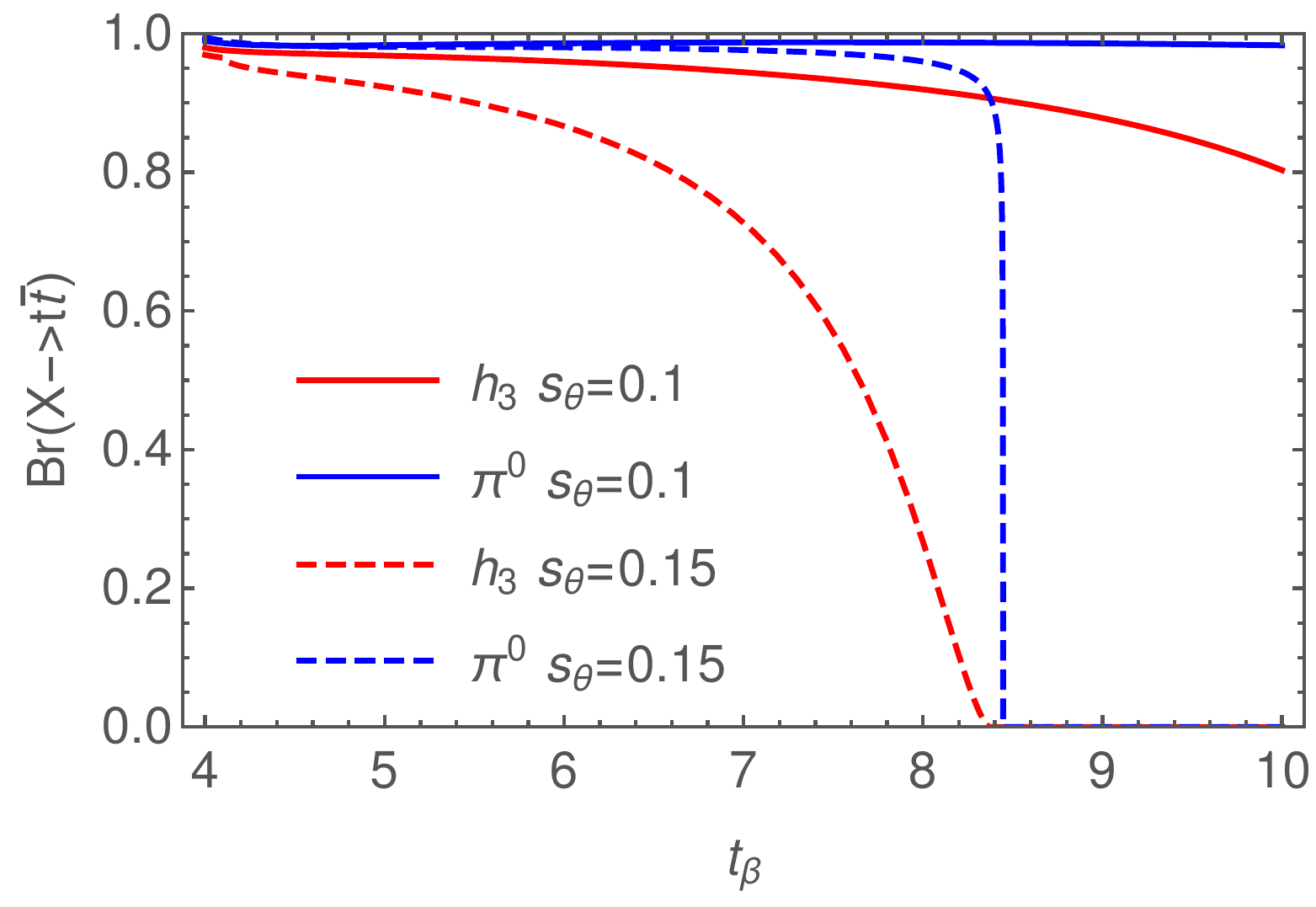} 
	 \caption{}
	\label{fig:h3pi3-Br}
	\end{subfigure}
	\caption{(a) Mass of the $\pi^0$. The dashed horizontal line shows the $t\bar{t}$ threshold. (b) Reduced coupling to top quarks.  
	    (c) Branching ratio of $h_3$ and $\pi^0$ to top-quark pair.
	    In all panels $\lambda_{\Phi}=0.15$ and $s_\theta=0.1\ (0.15)$ in the solid (dashed) lines.}
	\label{fig:h3pi3}
    \end{figure}
    On the other hand, the $\pi^0$ and $h_3$ states decay almost exclusively to $t\bar{t}$ 
    for $s_\theta\lesssim 0.1$ as can be seen in the branching ratio to $t\bar{t}$ shown in \fig{fig:h3pi3-Br} (off-shell decays 
    were neglected in the figure). At larger values of $s_\theta$, the masses can get lower than the $t\bar{t}$ threshold, $m_{\pi^0}<2m_t$, 
    and other decay channels become important, cf. Fig.~\ref{fig:h3pi3-mass}. In the case of $h_3$, the opening of other channels is 
    more severe since it couples at tree level to weak bosons and to di-Higgs. For the $\pi^0$ state, the $Zh$ channel becomes dominant.    
    
    In the case where the top decay channel prevails, we can have a situation in which the LHC might be able to observe these states.
    To see this, we first note that these states are nearly degenerate in mass (see \fig{fig:spectrum}) and therefore they would be seen 
    as a single summed peak at the LHC detectors. To access the magnitude of the peaks, we further note that the coupling of both states 
    to top quarks are very similar,
    and thus also their total widths (in the case of top dominated) as can be seen in \fig{fig:h3pi3-kt} (the variation with 
    $\lambda_\phi$ is very small). Therefore,  the production of both resonances can be effectively parametrized by the sum of the couplings. 
    The magnitude of this sum could be within the expected reach of LHC, even for heavy masses, if the systematic uncertainty in
    $t\bar{t}$ can be mildly improved~\cite{BuarqueFranzosi:2017qlm}.

\FloatBarrier

\section{Conclusions}
\label{sec:conclusions}

In this paper, we have studied the phenomenology of partially composite Higgs models based on the coset $\SU(4)/\Sp(4)$. In particular, 
we have considered the constraints on  
their parameter space from vacuum stability and perturbativity of the scalar self-couplings, 
together with constraints coming from Higgs coupling measurements, $B^0-\overline{B^0}$  mixing and LHC searches for new heavy scalars. 

We considered two 
realisations of the elementary scalar sector: a minimal one with a single EW 
doublet coupled to the new fermions, and one with an
$\SU(4)$ multiplet of scalars. In both models the new fermions feature vector-like current masses 
needed for the proper vacuum alignment; 
in the latter model they are dynamically generated by a vev of a EW-singlet component of the $\SU(4)$ scalar multiplet.

We have shown that
in the minimal model
there exists a region of allowed parameter values such that the vacuum remains stable 
and couplings perturbative up to the Planck scale. The EWSB is induced dynamically via the condensation in the strongly-interacting
new fermion sector. However, it is only possible to push the value of the mass parameter of the elementary scalar towards the 
compositeness scale $\Lambda = 4 \pi f$, and not beyond, at the expense of a lower vacuum stability scale. If we require stability of 
the vacuum up to the compositeness scale we find a bound on the mass parameter which is roughly ${m_H^2 \lesssim (0.1 \, s_\theta^{-1}\ \rm{TeV})^2}$.
Conversely, the parameter region with vacuum stability up to the Planck scale lies near the classically scale-invariant region.
However, we also show that LHC searches for new heavy CP-even scalars exclude this narrow stable window, while the constraints from 
a global fit to LHC Higgs coupling measurements and $B^0-\overline{B^0}$ mixing measurements are just beginning to probe this part of parameter space.

In summary, in most of the parameter space of the minimal model, the vacuum becomes unstable before the Planck
scale due to the enhanced top-Yukawa coupling of the elementary scalar as compared to the SM Higgs. This issue can be alleviated, if the 
top quark acquires a small part of its mass, e.g. of the order of the $b$-quark mass, from some other source than the elementary scalar. 
One such source could be ETC-type four-fermion operators. However, large modifications in the bottom-Higgs coupling would then lead to 
tension with the measurements in the stability region. 

The second model with a full $\SU(4)$ multiplet of scalars allows for 
dynamical generation of the vector-like masses of the new fermions, and the required explicit breaking of the $\SU(4)$ symmetry can 
be in the scalar potential as opposed to the Yukawa couplings in the minimal model.  
The quartic scalar couplings are in general enhanced compared to the SM removing the 
vacuum stability issue. However, due to the non-minimal scalar content, the singlet-scalar couplings instead develop Landau poles before 
the Planck scale in part of the parameter space. Again this limits the scalar mass parameters to the TeV region. One possibility to 
alleviate this problem is coupling the EW-singlet scalars to right-handed neutrinos.
This model is also less constrained by current LHC searches and $B^0-\overline{B^0}$ mixing, but 
the top decay channel might provide an interesting possibility to observe the heavier scalar states of the model at the 
LHC  in the future.

\section*{Acknowledgments}
We thank A. Kagan for discussions. 
TA acknowledges partial funding from a Villum foundation grant when part of this article was being completed.
 MTF acknowledges partial funding from The Council For Independent Research, grant number 
DFF 6108-00623. The CP3-Origins centre is partially funded by the Danish National Research Foundation, grant number DNRF90.
DBF acknowledges partial funding by the European Union as a part of the H2020 Marie Sk\l odowska-Curie Initial Training Network MCnetITN3
(722104).  

\clearpage

\appendix

\section{RG equations: elementary doublet}
\label{app:RG-D}
We consider a rather generic bTC framework featuring $\NF$ new 
$\SUL$-doublet fermionic fields, $Q_{\mathrm{L}}=(U_{\mathrm{L}},D_{\mathrm{L}})$, transforming under the representation $\RF$ under the new 
strong gauge group, and coupling to the elementary Higgs doublet via Yukawa interactions.  The one-loop evolution of the relevant couplings 
above the condensation scale, $\Lambda_{\mathrm{TC}}$, is given by
\begin{equation}
    \label{eq:}
    \begin{split}
	16\pi^2\beta_{g_{\mathrm{c}}}=&-7g_{\mathrm{c}}^3,\\
	16\pi^2\beta_{g_{\mathrm{TC}}}=&-\left(\frac{11}{3}C_2(A)-\frac{8}{3}\cdot \NF T(\RF)\right)g_{\mathrm{TC}}^3\, ,\\
	16\pi^2\beta_{\gL}=&-\left(\frac{19}{6}-\frac{2}{3}\NF d(\RF)T(\RF)\right)\gL^3,\\
	16\pi^2\beta_{\gY}=&\left(\frac{41}{6}+\frac{4}{3}\NF d(\RF)\left(4Y(Q_{\mathrm{L}})^2+\frac{1}{2}\right)\right)\gY^3,\\
	16\pi^2\beta_{y_t}=&\left(-8g_c^2-\frac{9}{4}\gL^2-\frac{17}{12}\gY^2+\frac{9}{2}y_t^2+d(\RF)(y_U^2+y_D^2)
	    \right)y_{t},\\
	16\pi^2\beta_{y_U}=&\left(-6C_2(\RF)g_{\mathrm{TC}}^2-\frac{9}{4}\gL^2-\frac{17}{12}\gY^2+3y_t^2+\left(d(\RF)
	    +\frac{3}{2}\right)y_U^2+d(\RF)y_D^2
	    \right)y_U,\\
	16\pi^2\beta_{y_D}=&\left(-6C_2(\RF)g_{\mathrm{TC}}^2-\frac{9}{4}\gL^2-\frac{5}{12}\gY^2+3y_t^2+\left(d(\RF)
	    +\frac{3}{2}\right)y_D^2+d(\RF)y_U^2
	    \right)y_D,\\
	16\pi^2\beta_{\lambda}=&24\lambda^2+\lambda\left(-3\left(3\gL^2+\gY^2\right)+12y_t^2+4d(\RF)\left(y_U^2+y_D^2
	    \right)\right)\\
	    &+\frac{3}{8}\left(2\gL^4+\left(\gL^2+\gY^2\right)^2)\right)-6y_t^4-2d(\RF)(y_U^4+y_D^4),
    \end{split}
\end{equation}
where 
\begin{equation}
    \label{eq:}
    \beta_g\equiv\frac{\mathrm{d}g}{\mathrm{d}\log\mu},
\end{equation}
$T(\RF)$ and $d(\RF)$ are the index and dimension of the representation $\RF$, and $C_2(A)$ is the quadratic Casimir of the 
adjoint representation. 

\section{RG equations: $\SU(4)$ multiplet of elementary scalars}
\label{app:RG-SU4}

Here we write down the RG equations for specific bTC framework based on $\SU(4)/\Sp(4)$ 
coset described in the main text. The new 
$\SUL$-doublet fermionic fields, $Q_{\mathrm{L}}=(U_{\mathrm{L}},D_{\mathrm{L}})$, transform in the fundamental representation of the 
new strong gauge group, 
and couple to the 
multiplet of elementary scalars transforming in the two-index antisymmetric representation of $\SU(4)$ via Yukawa interactions.  
From the RG point of view, the scalar sector corresponds to an EW doublet and two singlets, and the $\SU(4)$ invariance of the potential 
can be achieved only 
for one renormalisation scale $\mu_0$. The quartic scalar potential can be written as
\begin{equation}
    \label{eq:}
    V^{(4)}=\lambda_H(H^{\dagger}H)^2+\frac{\lambda_{S_R}}{4}S_R^4+\frac{\lambda_{S_I}}{4}S_I^4+\lambda_{HS_R}(H^{\dagger}H)S_R^2
	+\lambda_{HS_I}(H^{\dagger}H)S_I^2
	+\frac{\lambda_{S_RS_I}}{2}S_R^2S_I^2,
\end{equation}
with $\SU(4)$-symmetric boundary values $\lambda_H(\mu_0)=\lambda_{S_R}(\mu_0)=\lambda_{S_I}(\mu_0)=\lambda_{HS_R}(\mu_0)=\lambda_{HS_I}(\mu_0)
    =\lambda_{S_RS_I}(\mu_0)=\lambda_{\Phi}$.

The one-loop evolution of the relevant couplings above the condensation scale, $\Lambda_{\mathrm{TC}}$, is given by 
(we ignore the new Yukawa couplings, since for the viable parameter points able to give the correct Higgs mass, $y_Q\lesssim0.1$ throughout 
the parameter space we consider, and are thus their contribution to the $\beta$-functions is completely subdominant)
\begin{equation}
    \label{eq:}
    \begin{split}
	16\pi^2\beta_{g_{\mathrm{TC}}}=&-6g_{\mathrm{TC}}^3,\quad 16\pi^2\beta_{\gL}=-\frac{5}{2}\gL^3,\quad 16\pi^2\beta_{\gY}=\frac{15}{2}\gY^3,\\
	16\pi^2\beta_{y_t}=&\left(-8g_c^2-\frac{9}{4}\gL^2-\frac{17}{12}\gY^2+\frac{9}{2}y_t^2 
	    \right)y_{t},\\
	16\pi^2\beta_{\lambda_H}=&24\lambda_H^2+2(\lambda_{HS_R}^2+\lambda_{HS_I}^2)+\lambda_H\left(-3\left(3\gL^2+\gY^2\right)+12y_t^2
	    \right)\\
	    &+\frac{3}{8}\left(2\gL^4+\left(\gL^2+\gY^2\right)^2)\right)-6y_t^4, 
	    \\
	16\pi^2\beta_{\lambda_{HS_R}}=&8\lambda_{HS_R}^2+6\left(2\lambda_{H}+\lambda_{S_R}\right)\lambda_{HS_R}+2\lambda_{HS_I}\lambda_{S_RS_I}\\
	    &+\lambda_{HS_R}\left(-\frac{3}{2}\left(3\gL^2+\gY^2\right)+6y_t^2 
	    \right),\\
	16\pi^2\beta_{\lambda_{HS_I}}=&8\lambda_{HS_I}^2+6\left(2\lambda_{H}+\lambda_{S_I}\right)\lambda_{HS_I}+2\lambda_{HS_R}\lambda_{S_RS_I}\\
	    &+\lambda_{HS_I}\left(-\frac{3}{2}\left(3\gL^2+\gY^2\right)+6y_t^2 
	    \right),\\
	16\pi^2\beta_{\lambda_{S_R}}=&18\lambda_{S_R}^2+8\lambda_{HS_R}^2+2\lambda_{S_RS_I}^2, 
	\\
	16\pi^2\beta_{\lambda_{S_I}}=&18\lambda_{S_I}^2+8\lambda_{HS_I}^2+2\lambda_{S_RS_I}^2,\\
	16\pi^2\beta_{\lambda_{S_RS_I}}=&8\lambda_{S_RS_I}^2+6\lambda_{S_RS_I}(\lambda_{S_R}+\lambda_{S_I})+8\lambda_{HS_R}\lambda_{HS_I}.
    \end{split}
\end{equation}

%
\bibliography{PCH.bib}
\bibliographystyle{JHEP}
%
    
\end{document}